\RequirePackage{fix-cm}
\documentclass[natbib]{svjour3}                     
\smartqed  
\usepackage{graphicx}
\usepackage{epsf}
\usepackage{amssymb}
\usepackage{amsmath}
\usepackage{placeins}
\usepackage{color}

 \journalname{SSRv}
\newcommand{\be}{\begin{equation}}
\newcommand{\ee}{\end{equation}}
\newcommand{\beq}{\begin{eqnarray}}
\newcommand{\eeq}{\end{eqnarray}}
\newcommand\subsun[1]{{$_{\normalsize\odot}$}}

\def\lsim{\;\raise0.3ex\hbox{$<$\kern-0.75em\raise-1.1ex\hbox{$\sim$}}\;}
\def\gsim{\;\raise0.3ex\hbox{$>$\kern-0.75em\raise-1.1ex\hbox{$\sim$}}\;}

\begin{document}
\title{Relativistic Jets in Active Galactic Nuclei and Microquasars}

\titlerunning{Relativistic jets}        

\author{Gustavo E.~Romero \and
        M.~Boettcher \and
        S.~Markoff \and
        F.~Tavecchio
}


\institute{Gustavo E. Romero \at
              Instituto Argentino de Radioastronomía (IAR), C.C. No. 5, 1894, Buenos Aires, Argentina\\ 
              \email{romero@iar-conicet.gov.ar}
            \and
              M. Boettcher \at
              Centre for Space Research, Private Bag X6001, North-West University, Potchefstroom, 2520, South Africa\\
              \email{Markus.Bottcher@nwu.ac.za}
            \and
              S. Markoff \at
              Anton Pannekoek Institute for Astronomy, University of
              Amsterdam, 1098XH Amsterdam, The Netherlands \\
              \email{s.b.markoff@uva.nl}
            \and
           F. Tavecchio \at
             INAF-Osservatorio Astronomico di Brera, Via Bianchi 47, 23807 Merate, Italy\\
               \email{fabrizio.tavecchio@brera.inaf.it}
                }

\date{Received: date / Accepted: date}

\maketitle

\begin{abstract}
Collimated outflows (jets) appear to be a ubiquitous phenomenon associated with the accretion
of material onto a compact object. Despite this ubiquity, many fundamental physics aspects
of jets are still poorly understood and constrained. These include the mechanism of launching
and accelerating jets, the connection between these processes and the nature of the accretion
flow, and the role of magnetic fields; the physics responsible for the collimation of jets over 
tens of thousands to even millions of gravitational radii of the central accreting object; the 
matter content of jets; the location of the region(s) accelerating particles to TeV (possibly 
even PeV and EeV) energies (as evidenced by $\gamma$-ray emission observed from many jet sources) 
and the physical processes responsible for this particle acceleration; the radiative processes
giving rise to the observed multi-wavelength emission; and the topology of magnetic fields and
their role in the jet collimation and particle acceleration processes. This chapter reviews
the main knowns and unknowns in our current understanding of relativistic jets, in the context
of the main model ingredients for Galactic and extragalactic jet sources. It discusses aspects 
specific to active Galactic nuclei (especially blazars) and microquasars, and then presents a 
comparative discussion of similarities and differences between them. 

\keywords{Jets, outflows and bipolar flows \and Jets and bursts; galactic winds and fountains 
\and Active and peculiar galaxies and related systems \and X-ray binaries \and radiation mechanisms: non-thermal}
\end{abstract}

\section{Introduction}
\label{intro}

Relativistic jets are collimated outflows of plasma and fields
produced by accreting compact objects. The bulk motion of the plasma
is close to the speed of light, so relativistic effects play an
important role in the physical processes in these jets. The jet
phenomenon appears to be common wherever mass accretion onto a central
object occurs, and they become relativistic in the case of accretion
onto compact objects, such as neutron stars or black holes, both in
Galactic systems (neutron star and black-hole X-ray/$\gamma$-ray
binaries, hereafter termed generically XRBs --- termed microquasars 
when they exhibit prominent radio jets) and in extragalactic systems 
(active galactic nuclei [AGN], gamma-ray bursts [GRBs] and recently 
Tidal Disruption Events [TDEs]). Relativistic jets are ionized and 
produce non-thermal radiation from radio to gamma-rays in some cases. 
Observationally, in most cases, the presence of non-thermal radio 
emission has been the primary signature of relativistic jets, distinguishing 
jet-dominated radio-loud AGN from their radio-quiet counterparts (with very 
weak or absent jets), and microquasars (XRB states with prominent radio
jets) from radio-quiet XRB states. In this chapter, we will focus
on the relativistic jets from AGN and microquasars, while the physics
of GRB and TDE  jets is discussed extensively in chapters 6
-- 8, and jets from pulsars in chapter 2.

The existence of jet-like structures was known as a peculiarity of some galaxies such as M87 since the early twentieth 
century \citep{Curtis1918}. By the mid 1950s it was recognized that the extended radio emission in M87 was the result 
of synchrotron radiation \citep{Burbidge1956}. The synchrotron hypothesis was soon applied to explain the extended radio 
lobes observed in some galaxies \citep{Ginzburg1961,Shklovskii1961}. After the discovery of quasars \citep{Schmidt1963} 
at large cosmological redshifts, it became clear that some of the radio sources were extremely distant and the size and 
total energy output of the radio emitting clouds was huge. The cooling time of the electrons responsible for the 
radiation is much shorter than the dynamical timescales, thus
indicating a need for in-situ acceleration of relativistic particles
along the jets and in the radio lobes. Fully non-thermal radio jets were soon detected in many radio galaxies 
using interferometric techniques. Structural time variations were also detected in these jets, and superluminal 
motions, originally predicted by \cite{Rees1966} were found, first in the jet of the quasar 3C273, and later in 
many other sources \citep[see][and references therein]{Zensus1987}.

In our own galaxy, relativistic jets were associated with the emission
line star SS433 \citep[e.g.,][]{Abell1979}.  The jets of this source,
a compact object accreting from a companion star, precess with a
164-day period. Only much later, in the 1990s, other X-ray binary
sources mimicking the behaviour of extragalactic quasars were found
\citep{Mirabel1992} and superluminal motions observed within the
Galaxy \citep{Mirabel1994}.  XRBs with relativistic radio jets were
dubbed ``microquasars''.  About 15 such sources are currently
classified as microquasars because of explicit measurements of
relativistic proper motion, but $\sim$50 XRBs have by now been
observed in the jet-dominated 'hard state' \citep[][and see
Section~\ref{microquasars}]{Tetarenkoetal2016}, and many more are
suspected to exist in the Milky Way.

With the advent of orbiting X-ray observatories such as {\it Einstein},  {\it EXOSAT}, {\it ROSAT}, and more 
recently {\it RXTE}, {\it XMM-Newton}, {\it Chandra}, {\it Swift}, and {\it NuSTAR}, and the opening-up of the $\gamma$-ray 
sky especially with the Compton Gamma-Ray Observatory in the 1990s and
more recently with {\it AGILE}] and the 
{\it Fermi} Gamma-Ray Space Telescope, it became soon clear that both Galactic and extragalactic 
jets are high-energy sources. Their spectral energy distribution (SED) is strongly dominated by non-thermal 
emission from radio up to $\gamma$-ray frequencies, and in many cases it is dominated by the $\gamma$-ray 
component that extends up to GeV and even TeV energies \citep[see, e.g.,][]{MaraschiTavecchio2001,Abdoetal2010}. 
The emission from these objects tends to be highly variable at all wavelengths, with often complex --- sometimes 
correlated, but sometimes also uncorrelated among different wavelength regimes --- flux and spectral variability 
patterns, that may change character even within the same source among different eposides of activity. The study 
of these complex radiative features of relativistic jets is the only way to gain insight into the physical 
conditions of the outflowing plasma and the mechanisms that result in the jet launching and subsequent particle 
acceleration. However, even after over three decades of intensive multi-wavelength observing campaigns (including 
observations from the radio to $\gamma$-rays) and remarkable progress
in general relativistic magnetohydrodynamic (GRMHD) and particle-in-cell 
(PIC) simulations of plasma phenomena relevant to jet launching, the dynamics, and particle acceleration, many 
fundamental aspects of the jet phenomenon are still very poorly
understood or even constrained.  This 
chapter attempts to summarize the cutting edge, as well as elucidate potential avenues for future research,
in order to gain insight into the answers to these 
open questions:

\begin{itemize}

\item What is the primary physical mechanism powering the acceleration
  and collimation of the jet? Is the primary energy source the
  rotational energy of the black hole \citep{BlandfordZnajek1977}, or
  is power released in the accretion flow the primary driver
  \citep{BlandfordPayne1982}? In the latter case, one would expect a
  tight connection between the observational signatures of the
  accretion flow and the presence/absence of relativistic jets, as is,
  indeed, observed in many Galactic systems, but also evidence for a
  correlation between black-hole spin and jet power has been claimed
  \citep[e.g.,][see Section~\ref{microquasars} for a discussion]{McClintock2014}. 
  In general, what is the role of magnetic fields in the acceleration 
  and collimation of relativistic jets?

\item To what extent can the accretion and jet processes involving supermassive black holes in AGN be considered 
as simply scaled-up versions of the microquasar phenomenon? While the source of the accreted material and thus
the accretion-flow geometry is quite different, many aspects, such as mass-scaled short-term variability 
patterns and the characteristic two-component, non-thermal spectral energy distributions, appear to indicate 
a large degree of analogy between the two phenomena on different scales \citep[e.g.,][]{Markoff2015}. 

\item What is the strength and topology of magnetic fields along relativistic jets, and what role do they
play in the collimation of jets out to tens of thousands of
gravitational radii, as well as the acceleration of relativistic
particles in-situ along the jets?  What determines where particle
acceleration begins,  a region often modeled as a single zone for
e.g., blazars (see Section~\ref{blazars}), but also becoming a focus
now for XRBs (see Section~\ref{microquasars}).  

\item What is the matter content of material flowing along the jets? Is it primarily an electron-positron
pair plasma, or an ordinary electron-proton plasma \citep[see, e.g.,][]{SikoraMadejski2000}? If protons 
are present in significant numbers, are they also subject to relativistic particle acceleration?  

\item What are the dominant radiation mechanisms leading to the
  observed two-component non-thermal emission seen from relativistic
  jet sources? Are these primarily leptonic processes, with protons
  (if present) not contributing to the radiative output, or are 
  protons also accelerated to ultrarelativistic energies so that hadronic
  processes can make a significant contribution \citep[see, e.g.,][for
  a comparative analysis of both types of models in the case of
  blazars]{Boettcheretal2013}? If the latter is the case, then AGN may
  be the elusive sources of ultra-high-energy cosmic rays (UHECRs,
  i.e., cosmic-rays with energies $E > 10^{19}$~eV). Also, the
  acceleration of UHECRs in AGN environments is likely to result in
  charged-pion production, leading to the emission of PeV neutrinos,
  as recently confidently detected by the IceCube neutrino observatory
  \citep{Aartsen2014}.

\end{itemize}

In this chapter, we will discuss the physics and phenomenology that is
common to relativistic jets on different size scales, from
astronomical units to kiloparsecs. We will focus on the jets of
microquasars and blazars, with emphasis on their radiative output. We
will first introduce the basic, common model ingredients required for
any model of relativistic jets of different sizes and velocities
(Section \ref{model}). Against this background, we will offer an
overview of the current knowledge about particle acceleration within
the jets of blazars (Section \ref{blazars}) and microquasars (Section
\ref{microquasars}). The final section (\ref{discussion}) provides a
comparative discussion of the prevalent issues in blazar and
microquasar jet physics, and an outlook towards future experiments
that may shed light on these issues. The reader interested in
extensive treatments of this subject may consult the recent
monagraphs and collections by \cite{Beskin2009}, \cite{Belloni2010},
\cite{Boettcher2010}, \cite{Markoff2010} and \cite{RomeroVila2014}.

\section{Basic Model Ingredients: Knowns and (Educated) Guesses}
\label{model}

Wherever accretion of matter with angular momentum and magnetic fields
into a gravitational well occurs, an accretion disk is formed, often
accompanied by the expulsion of collimated outflows. When the
gravitating object is compact (a weakly magnetized neutron star or a
black hole), it is capable of launching relativistic outflows,
although the exact mechanism is not yet clear. Models of relativistic jets
were first developed for Active Galactic Nuclei (AGN) at the end of
the 1970s and early 1980s
\citep[e.g.,][]{BlandfordKoenigl1979,Marscher1980a,Marscher1980b,MarscherGear1985}
and later adapted to microquasars (by then not known with this name)
by
\cite{BandGrindlay1986,HjellmingJohnston1988,LevinsonBlandford1996}.

In these early models for the broadband emission from jets, it is
assumed that only leptons (electrons and positrons) are accelerated,
and thus contributing significantly to the radiation, with protons (if
present) not achieving sufficiently high energy. Such models are
termed {\it leptonic models}, and they have achieved an increasing
degree of sophistication over the past several decades, both for
Galactic
\citep[e.g.,][]{Markoffetal2001,Markoffetal2005,Georganopoulosetal2002,Kaufmanetal2002,BoschRamonetal2006,Zdziarskietal2014}
and extragalactic jets
\citep[e.g.,][]{Maraschietal1992,DermerSchlickeiser1993,BlandfordLevinson1995,LevinsonBlandford1995,Markowithetal1995,
  Romeroetal1995,Boettcheretal1997,BoettcherChiang2002,Markoff2015}.

Alternatively, if protons are accelerated to ultrarelativistic
energies they may contribute significantly to the radiative output,
and the corresponding class of models is termed {\it hadronic models}
or {\it lepto-hadronic models}, to indicate that even in those models,
relativistic leptons make a significant contribution to the SED,
especially at low frequencies. For lepto-hadronic jet models for
Galactic sources, see, e.g.,
\cite{Romeroetal2003,Romeroetal2005,RomeroVila2008,VilaRomero2010,Vilaetal2012,YanZhang2015,Cerrutietal2015,Pepeetal2015}
and for extragalactic sources,
\cite{Mannheim1993,RachenBiermann1993,Muckeetal2003,Reynosoetal2011,Reynosoetal2012,Boettcheretal2013}.

As indicated in Section \ref{intro}, many aspects of the jet geometry, magnetization, mode of particle acceleration
and the location of the dominant energy dissipation and particle acceleration zone are currently still unknown,
leaving a significant amount of freedom in the choice of model parameters. 
The various models of relativistic jets differ in the characteristics of the injection of relativistic particles, 
the properties of the region in the jet where the radiation is produced, the particles and physical interactions 
involved in the production of the radiation, the external conditions, the treatment of the transport of particles 
in the radiative zone, the inclusion of dynamical effects, the boundary and external medium conditions, the 
radiation reprocessing in the source, and many other factors. 

The emission of high-energy and very-high-energy $\gamma$-rays
requires the acceleration of particles to GeV and even TeV energies,
which is likely to occur in relatively small, localized regions along
the jet.  Due to their short cooling time scale, those particles are
expected to emit high-energy radiation locally.  The concept of a
single, dominant, small emission region is further strengthened by the
observed rapid variability which, due to the causality constraint,
restricts the size of the emission region, in many cases, to the order
of the gravitational radius of the central accreting object
\citep[e.g.,][]{Begelmanetal2008}.  This gives credence to single-zone
models, where the emitting region is considered to be homogeneous
\citep[usually assumed spherical but see, e.g.][]{GianniosUzdenskyBegelaman2009} 
and the physical conditions uniform. Such models have been popular because of 
their relative simplicity.  However, the often observed lack of strict 
correlations between variability patterns in different wavelength regimes
demonstrates that they may be applicable primarily only to short-term,
isolated flaring events in which the emission is likely to be
dominated by a single, small emission region.  Nevertheless,
single-zone models may be characterized by a limited number of
parameters, many of which can be reasonably well constrained from
observables (for details on such observational parameter constraints,
see Section \ref{blazars} and \ref{microquasars}). Hence, they have
resulted in significant insight into the physical conditions and
physical processes in the dominant high-energy emission region.

The necessity of inhomogenous (multi-zone) models has already been
pointed out, e.g., by \cite{Ghisellinietal1985}, but their
implementation in a self-consistent way has been hampered by the large
number of (poorly constrained) parameters required for such
models.    Only recently inhomogenous models with an adequate treatment
of particle transport and all radiative processes have been developed.
However these models have still not been made self-consistent with the
magnetohydrodynamics (MHD) governing the plasma properties and flow
characteristics, and which may determine how structures that
accelerate particles are created.   Ultimately the number of free 
parameters will be reduced once the macro- and microphysics can 
be linked, but this type of work is still in its early stages.  
 
\begin{figure}%
\centering
\includegraphics[width=0.7\columnwidth, keepaspectratio, trim= 0 90 350 50, clip]{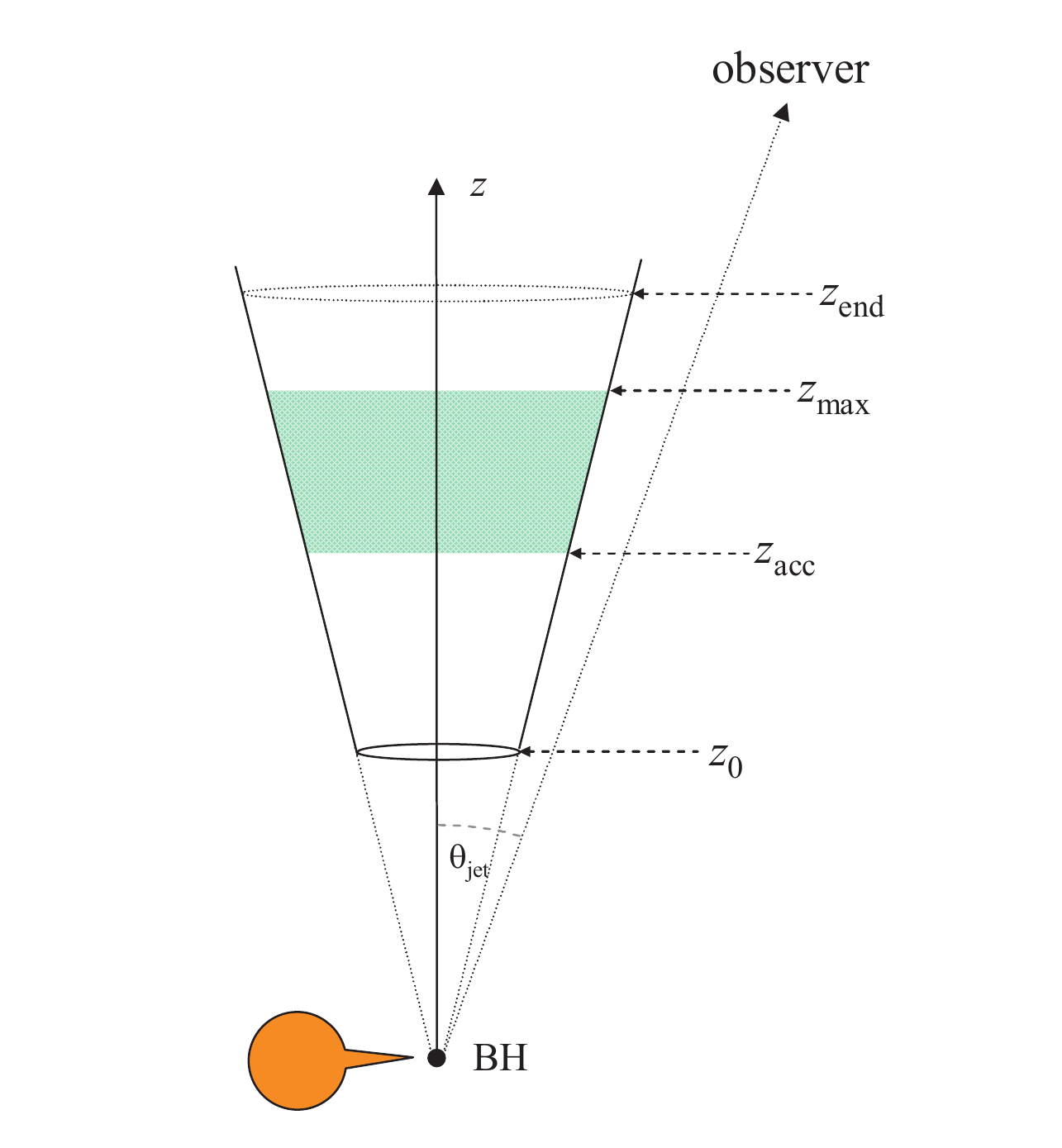}%
\caption{Detail of a generic conical jet model, including the acceleration region (not to scale). Some 
relevant parameters are indicated.}%
\label{fig:jet_detail}%
\end{figure}

In what follows we will discuss the most important physical processes
that enter relativistic jet models (both leptonic and lepto-hadronic) 
for both Galactic and extragalactic sources 
in a generic model geometry as illustrated in Figure \ref{fig:jet_detail}.
The discussion presented 
here rests upon the work by
\cite{MarkoffNowakWilms2005,Vilaetal2012,Pepeetal2015,DiltzBoettcher2014,Reynosoetal2012,PolkoMeierMarkoff2014,Diltzetal2015,Zhangetal2014,Zhangetal2015}.

\subsection{Energetics and geometry}
\label{sec:energetics}

A compact object (hereafter assumed to be a black hole) of mass $M_{\rm{BH}}$ accretes matter from its environment 
(in the case of a microquasars the accretion flow comes from a companion star).  We write the accretion power 
$L_{\rm{accr}}$ in terms of the Eddington luminosity of the black hole as

\begin{equation}
L_{\rm{accr}}\equiv \dot{M}c^2 = q_{\rm{accr}}\,L_{\rm{Edd}}\approx 1.3\times10^{46} q_{\rm{accr}}
\left(\frac{M_{\rm{BH}}}{10^8\;M_{\odot}}\right) \mathrm{erg  \,s}^{-1},
\label{eq:accretion_power}
\end{equation}

\vspace{0.2cm} 

\noindent where $\dot{M}$ is the mass accretion rate, $M_{\odot}$ is
the solar mass, and $q_{\rm{accr}}$ is a dimensionless parameter. The
accretion disk extends from an inner radius $R_{\rm{in}}$ to an outer
radius $R_{\rm{out}}$.  For particularly lower luminosity
(sub-Eddington) sources, within $R_{\rm{in}}$ the plasma is thought to
become radiatively inefficient, ``inflating'' to form a hot, optically
thin corona.   The nature and geometry of this corona is a matter of
significant debate \citep[see, e.g.][]{Nowaketal2011}, however the
existence of a population of hot electrons is necessary in any
scenario.

A fraction of the accreted matter is ejected into two symmetrical jets, each carrying a power

\begin{equation}
L_{\rm{jet}}= \frac{1}{2}q_{\rm{jet}}\,L_{\rm{accr}},
\label{eq:jet_power}
\end{equation}

\vspace{0.2cm} 

\noindent where $q_{\rm{jet}}<1$.  Note however that this assumption
implicitly assumes the \cite{BlandfordPayne1982} mechanism, thus if
enhancement from the black hole spin is relevant
\citep{BlandfordZnajek1977} this linear relationship with the
accretion disk luminosity may no longer hold.  

The jets are launched perpendicularly to the plane of the inner accretion
disk at a distance $z_0$ from the black hole. The scale $z_0$ is the 
characteristic scale of the coronal plasma that is thought to surround the 
black hole \citep{Romeroetal2010}. The jet might be the result of the evacuation 
of the corona by magnetocentrifugal effects, a hypothesis supported by the 
fact that jets in XRBs are observed only in the hard state, when the accretion 
disk is trunctated at a distance of the order of $z_0$. A spherical geometry for 
the corona seems to be the more appropriate on the basis of the intensity of the 
Compton reflection feature observed in some sources \cite[e.g.][]{Poutanenetal1997}. 
It has been also suggested that the corona is essentially identical to the base of 
the jet \citep{Markoffetal2005}. In any case, the full formation of the collimated 
outflow takes place at a distance $z_0$, and from there on the jet can be treated 
as a distinct physical component of the system. 

The jet axis makes an
angle $\theta_{\rm{jet}}$ with the line of sight to an observer on
Earth. The exact geometry of the jet will be determined by the balance
of internal and external pressure, and thus strongly depends on the
plasma properties at launch, which are currently undetermined.  There
is evidence, e.g. in the case of M87, that the jet base is described
by an approximately paraboidal shape, transitioning to a conical
structure \citep{AsadaNakamura2012,PotterCotter2013} consisistent with
predictions of a collimated, magnetically-dominated flow that is
converted to a kinetically dominated flow.  At large distances
($z \gg 10^5 \, r_g$, where $r_g = G M_{\rm BH} / c^2$ is the
gravitational radius of the black hole), jets appear to be well
collimated in a quasi-cylindrical structure.  The simplest assumption,
often adopted in the literature, is thus a constant jet radius over
the section of the jet responsible for the majority of the radiative
output. Alternatively, a conical geometry is often assumed, so that
the jet cross-sectional radius grows as

\begin{equation}
r(z)=r_0\left(\frac{z}{z_0}\right).
\label{eq:jet_radius}
\end{equation}

\noindent 
If magnetic energy is the primary driver of the outflow (i.e., the
plasma is ejected by some magnetocentrifugal mechanism or driven by
Poynting flux), the magnetic energy density at the base of the jet
must be enough to set the plasma in motion. This requires at least as
much energy in the magnetic field as kinetic energy the jet will
achieve. In this case, the value $B_0$ of the magnetic field at $z_0$
may be constrained by the condition of equipartition between the
magnetic and the kinetic energy densities,
$U_{\rm{mag}}(z_{\rm{0}}) \ge U_{\rm{kin}}(z_{\rm{0}})$.  Then,
assuming magnetic-flux conservation, the magnetic energy density
decreases as the jet expands. One may parameterize the dependence on
$z$ of the magnetic field strength as

\begin{equation}
B(z)= B_0\left(\frac{z_0}{z}\right)^m,
\label{eq:magnetic_field}
\end{equation}
 
\noindent where the index $m \geq 1$ depends on the magnetic-field topology \citep[e.g.][]{Krolik1999}. In the case of
a conical jet and a purely poloidal magnetic field (i.e., $B = B_z$), $m = 2$, whereas for a purely toroidal B-field 
(i.e., $B = B_{\phi}$), $m = 1$. 

However, in the case of an MHD-driven outflow, such conservation is
likely not valid. In a widely accepted model of jet acceleration \citep[see, for
example,][]{Spruit2010}, a fraction of the magnetic energy is
converted into kinetic energy of the plasma. The bulk Lorentz factor
of the jet, $\Gamma_{\rm{jet}}$, then increases up to a maximal
value. The behavior of $\Gamma_{\rm{jet}}$ with the distance to the
black hole has been studied both analytically and numerically in the
context of ideal MHD
\citep[e.g.,][]{Lyubarsky2009,Tchekhovskoyetal2008,Tchekhovskoyetal2010,
  PolkoMeierMarkoff2014}.

In the context of this simple model we parameterize dissipation in the region
\mbox{$z_{\rm{acc}}\leq z \leq z_{\rm{max}}$}, where some fraction of the
jet power is transformed into kinetic energy of relativistic electrons
(and protons in the case of lepto-hadronic models). Possible
acceleration mechanisms include diffusive shock acceleration
\citep[DSA, see, e.g.,][]{SummerlinBaring2012} or, in particular for
jets with high magnetization, fast magnetic reconnection in presence
of turbulence \citep[e.g.,][]{Lazarianetal2015}. If DSA is the
dominant acceleration mechanism, the value of $z_{\rm{acc}}$ is
constrained by the fact that, for shock waves to develop, the magnetic
energy density of the plasma must be in sub-equipartition with the
kinetic energy density \citep{Komissarovetal2007}. The power
transferred to the relativistic particles is parameterized as

\begin{equation}
L_{\rm{rel}}= q_{\rm{rel} }L_{\rm{jet}},
\label{eq:power_rel_particles}
\end{equation}

\noindent where $q_{\rm{rel}}<1$ . The value of  $L_{\rm{rel}}$ is the sum of the power injected in both protons 
and electrons 

\begin{equation}
L_{\rm{rel}} = L_p + L_e \equiv (1 + a) \, L_e
\label{eq:power_rel_p_e}
\end{equation}

\noindent where we have defined an energy partion parameter $a$ such that $L_p  = a \,L_e$. If $a > 1$ the jet is 
proton-dominated
(i.e., hadronic or lepto-hadronic models), whereas for $a<1$ most of the power is injected in relativistic 
electrons (i.e.,
leptonic models).

\subsection{Microphysical particle treatment}
\label{sec:microphysics}

The relativistic particles in the jet lose energy by radiation and also adiabatic losses in the case of an expanding jet. 
Several processes contribute to the radiative cooling, since particles can interact with the magnetic, radiation, and 
matter fields of the jet.

The most important interaction channel with the magnetic field of the jet is synchrotron radiation. For a charged 
particle of mass $m$, energy $E = \gamma m c^2$, and charge $q = Z \, e$, the synchrotron energy-loss rate is 
\citep[e.g.,][]{BlumenthalGould1970}

\begin{equation}
\left.\frac{d\gamma}{dt}\right|_{\rm{synchr}} = -  \frac{4}{3}\left(\frac{m_e}{m}\right)^3\frac{c\sigma_{\rm{T}}\, 
U_{\rm{mag}}}{m_ec^2} \, Z^4 \, \gamma^2,
\label{eq:loss_rate_synchr}
\end{equation}

\vspace{0.2cm} 

\noindent where $c$ is the speed of light, $m_e$ is the electron mass, $\sigma_{\rm{T}}$ is the Thomson cross-section, 
and $U_{\rm{mag}}$ is the magnetic energy density. The ratio $\left(m_e/m\right)^3$ in Eq. (\ref{eq:loss_rate_synchr}) 
makes synchrotron cooling particularly efficient for the lightest particles.
 
Relativistic electrons also cool through inverse Compton (IC)
scattering off the jet and potentially also external radiation fields. Writing the characteristic
photon energy of the target photon field as
$\epsilon = h \nu / (m_e c^2)$, the Compton scattering process with
electrons is elastic in the electron rest frame if
$\epsilon\gamma \ll 1$, which denotes the Thomson regime. In this
regime, the Compton cooling rate is identical to Equation
\ref{eq:loss_rate_synchr}, except for replacing the energy density
$U_{rm mag}$ by the energy density of the target radiation field,
$U_{\rm rad}$.  Due to the same suppression factor $(m_e / m)^3$,
Compton scattering is very inefficient for particles heavier than
electrons/positrons, and is therefore usually considered only for
leptons. Only in cases where extreme energies for protons are
possible, this channel must be taken into account also for heavier
particles \citep[e.g.,][]{Aharonian2000,Aharonian2002}). For the more
general case, in which the Thomson condition is not fulfilled (i.e.,
including the Klein-Nishina regime with $\epsilon\gamma \gtrsim 1$),
the energy-loss rate needs to be evaluated considering recoil in the
rest frame,  using the full Klein-Nishina cross section,
e.g. following \cite{BlumenthalGould1970}; see
\cite{Boettcheretal1997,DermerMenon2009,Boettcheretal2012} for
detailed calculations of the Compton loss rates specific to blazars.

An important and ubiquitous target photon field is the synchrotron radiation co-spatially produced within the
jet. This is referred to as the synchrotron-self-Compton (SSC) process, which often dominates the X-ray (and sometimes
also $\gamma$-ray) emission in relativistic jet models. In a homogeneous single zone of radius $R$ in steady-state,
the radiation energy density $U_{\rm rad}$ is related to the emissivity $j_{\rm synchr}$ (i.e., the energy production 
rate per unit volume in synchrotron radiation) by 

\begin{equation}
U_{\rm synchr}^{\rm homog} = {3 \over 4} \, {R \over c} \, j
\label{Urad_homogeneous}
\end{equation}
In more complex geometries, a convenient treatment lies in the ``local approximation'' of \cite{Ghisellinietal1985}:

\begin{equation}
U_{\rm{synchr}}^{\rm local} (z) \approx j_{\rm{synchr}}(z)\,\frac{r_{\rm{jet}}(z)}{c},
	\label{eq:sy_local}
\end{equation}  

Additional target photon fields for Compton scattering can originate outside the jet and include the radiation field
from the accretion disk and of the stellar companion, in the case of microquasars, as well as line-dominated radiation
from the Broad Line Region, Narrow Line Region, and a dusty torus around the central accretion flow in the
case of AGN (see Sections \ref{blazars} and \ref{microquasars} for more details). 

The interaction of relativistic protons with radiation can create electron-positron pairs (``photopair'' production
via the Bethe-Heitler process)  

\begin{equation}
p+\gamma\rightarrow p+e^-+e^+.
	\label{photopair}
\end{equation}  

\noindent The photon threshold energy for this process is $\sim 1$ MeV in the rest frame of the proton. At higher energies, 
proton-photon collisions can also yield pions (``photomeson'' production). The two main channels are

\begin{equation}
p+\gamma\rightarrow p+a\pi^0+b\left(\pi^++\pi^-\right)
	\label{eq:photomeson1}
\end{equation}  

\noindent and
 
\begin{equation} 
p+\gamma\rightarrow n+\pi^++a\pi^0+b\left(\pi^++\pi^-\right),
	\label{eq:photomeson2}
\end{equation}  

\noindent where the integers $a$ and $b$ are the pion multiplicities. Both channels have approximately the same 
cross-section, and a photon threshold energy of $\sim145$ MeV in the proton rest-frame, corresponding to a proton
energy, in the laboratory frame, of

\begin{equation}
E_p^{\rm thr} = {m_p c^2 \, m_{\pi} c^2 \over 2 \, E_{\rm ph}} \, \left( 1 + {m_\pi \over 2 \, m_p} \right) \sim
7 \times 10^{16} \, E_{\rm eV}^{-1} \; {\rm eV}
\label{pionthreshold}
\end{equation} 
where $E_{\rm eV}$ is the characteristic photon energy of the target photon field in units of eV. The subsequent 
decay of charged pions injects leptons and neutrinos

\begin{equation}
\pi^+\rightarrow\mu^++\nu_\mu, \quad \mu^+\rightarrow e^++\nu_{\rm{e}}+\overline{\nu}_\mu,
	\label{eq:piondecay1}	
\end{equation}

\begin{equation}
\pi^-\rightarrow\mu^-+\overline{\nu}_\mu, \quad \mu^-\rightarrow e^-+\overline{\nu}_{\rm{e}}+\nu_\mu,
	\label{eq:piondecay2}	
\end{equation}

\noindent whereas neutral pions decay into two photons,

\begin{equation}
\pi^0\rightarrow\gamma+\gamma.
	\label{eq:piondecay3}	
\end{equation}

For photomeson and photopair production, the proton energy-loss rate can be calculated as in \cite{Begelmanetal1990}. 
Hadronic / lepto-hadronic models involving photo-meson production require protons to be accelerated to energies of 
at least the threshold energy given in Equation (\ref{pionthreshold}). Confining protons of this energy to the small
emission region sizes inferred from rapid variability, typically requires magnetic fields of $B \gtrsim 10$ -- $100$~G.
In the presence of such strong magnetic fields, the electron-synchrotron radiation field is expected to be strongly
dominant over any potential external radiation fields. Therefore, one typically only considers the synchrotron photons 
of primary electrons as the target photon field for photo-pair and photo-pion production. $\gamma$-rays are produced
through the $p\gamma$ interaction channels primarily by synchrotron radiation of charged secondaries 
($e^{\pm}$, $\mu^{\pm}$, $\pi^{\pm}$), and the decay of neutral pions. 

High-energy protons can also interact with thermal particles in the jet plasma. Due to the low densities present
in extragalactic jet sources, proton-proton interactions are usually negligible in AGN jets, but they can make a significant 
contribution to the proton loss rate and radiation output in the case of the much denser Galactic jet sources. If the 
energy of the relativistic proton is higher than the threshold value for $\pi^0$ production ($\approx 1.22$~GeV), the 
collision with a thermal proton can create pions
\begin{equation}
p+p\rightarrow p+p+a\pi^0+b\left(\pi^++\pi^-\right),
	\label{pp1}
\end{equation}  

\begin{equation}
p+p\rightarrow p+n+\pi^++a\pi^0+b\left(\pi^++\pi^-\right).
	\label{pp2}
\end{equation}    

\noindent As in proton-photon interactions, proton-proton collisions inject photons by means of the decay of neutral 
pions. Charged pions inject electron-positron pairs and neutrinos through the decay chains of Eqs. (\ref{eq:piondecay1}) 
and (\ref{eq:piondecay2}). 

The energy-loss rate for a proton of energy $E$ is given by \citep[e.g.,][]{Begelmanetal1990}

\begin{equation}
\left.\frac{dE}{dt}\right|_{pp} = -n_p\,c\,E\,\kappa_{pp}\sigma_{pp},
	\label{eq:pp_loss_rate}
\end{equation}    

\noindent where $n_p$ is the number density of thermal protons in the co-moving frame\footnote{Henceforth, 
we refer to as ``comoving'' or ``jet frame''  the reference frame fixed to the jet plasma, where the outflow 
bulk velocity is zero. The ``observer frame'' corresponds to the frame in which the jet bulk velocity is 
$v_{\rm{jet}}$.} , $\sigma_{pp}$ is the proton-proton cross-section, and \mbox{$\kappa_{pp}\approx0.5$} is 
the total inelasticity of the interaction. A convenient parametrization for $\sigma_{pp}$ is presented in 
\cite{Kelneretal2006}.  Taking the assumptions about distribution of
jet power above, we can calculate the value of $n_p$ in the comoving
frame \citep[see also][]{BoschRamonetal2006}:

\begin{equation}
n_{p}=\left(1 - q_{\rm{rel}}\right)\frac{L_{\rm{jet}}}{\Gamma_{\rm{jet}}^2 \pi r_{\rm{jet}}^2v_{\rm{jet}}m_pc^2},
	\label{eq:thermal_particles_density}  
\end{equation} 

\noindent where $m_p$ is the proton mass.

Finally, in the case of an expanding jet, particles also lose energy due to adiabatic cooling. The corresponding
adiabatic energy loss rate is given by

\begin{equation}
\left.\frac{dE}{dt}\right|_{\rm{ad}}(E,z) = -\frac{E}{3}\frac{d V}{d t}.
\label{eq:adiab_loss_rate}
\end{equation}

\noindent which obviously depends on the jet geometry \citep[see, e.g.,][]{BoschRamonetal2006}.

\noindent The total energy-loss rate is simply the sum of the relevant loss rates for the particle species 
in consideration

\begin{equation}
\left.\frac{dE}{dt}\right|_{\rm{tot}}(E,z) = \sum_i{\left.\frac{dE}{dt}\right|_{i}(E,z)}.
\label{eq:total_loss_rate}
\end{equation}

For particles whose gyro-radii are smaller than the acceleration / radiation region
(this is typically always the case for electrons/positrons),
the maximum energy of primary electrons and protons can be estimated by equating the total 
energy-loss rate to the acceleration rate

\begin{equation}
-\left.\frac{dE}{dt}\right|_{\rm{tot}}(E_{\rm{max}}) = \left.\frac{dE}{dt}\right|_{\rm{acc}}(E_{\rm{max}}). 
\label{eq:max_energy}
\end{equation}

However, for protons, the confinement condition (i.e., the gyro-radius equalling the size
of the acceleration region) is sometimes the dominant factor limiting the maximum energy, rather
than radiative cooling.
The magnitude and energy dependence of the particle acceleration rate obviously depends on the dominant
acceleration mechanisms. However, it is often convenient to parameterize diffusive shock acceleration 
processes through an acceleration efficiency parameter $\eta < 1$ so that the acceleration time scale 
is a multiple $\eta^{-1} > 1$ of the gyro-timescale. Consequently, the acceleration rate is then given 
by  \citep[e.g.,][]{Aharonian2004}. 

\begin{equation}
	\left.\frac{dE}{dt}\right|_{\rm{acc}} = \eta\,e\,c\,B(z),
	\label{eq:acc-rate}
\end{equation}

Pions and muons can also interact before decaying. This effect becomes important when particles are very 
energetic and the physical conditions in the jet are such that the cooling time is shorter than the mean 
lifetime \citep{ReynosoRomero2009}. Pion and muon synchrotron radiation (and cooling) may be neglected if
the maximum primary proton Lorentz factor and the magnetic field satisfy the conditions  \citep{Boettcheretal2013}:

\begin{equation}
B \, \gamma_p \ll \begin{cases}{7.8 \times 10^{11} \; {\rm G}} & {\rm for \; pions} \\
                               {5.6 \times 10^{10} \; {\rm G}} & {\rm for \; muons}
                   \end{cases}
\label{Bgammap}
\end{equation}

The injection of secondary particles is not a consequence of diffusive shock acceleration, but a by-product 
of the interaction of primary protons and electrons.  The maximum energy of pions, muons, and secondary 
electron-positron pairs is then not fixed by Eq. (\ref{eq:max_energy}). It is instead determined by the 
characteristics of the particular process through which the different particles were injected, and by the 
maximum energy of the primary particles (which is constrained by Eq. \ref{eq:max_energy}). 

The injection of primary particles into the emission region may represent either the entrainment of external
material by the flow, or a very rapid (e.g., first-order Fermi) acceleration process, on a time scale much
shorter than all other relevant time scales. Such injection is thus usually trated as a phenomenological
source term $Q(E, z)$ in the co-moving jet frame, which is often parameterized either as a $\delta$ function
in particle energy, 

\begin{equation}
Q(E, z)\Bigr\vert_{\rm entrainment} = Q_0 \, \delta (E - \Gamma_{\rm jet} \, m c^2)
\label{Qentrainment}
\end{equation}
in the case of entrainment by the relativistic flow of Lorentz factor $\Gamma_{\rm jet}$ 
\citep[e.g.,][]{PohlSchlickeiser2000}, or as a power-law in energy multiplied by an exponential 
cutoff,

\begin{equation}
Q(E,z) = Q_0\,E^{-p}\exp\left[-E/E_{\rm{max}}(z)\right]\,f(z).
\label{eq:injection_function}
\end{equation}
 
\noindent in the case of diffusive shock acceleration. In the latter case, the value of the spectral index is 
typically in the range $1.5 \lesssim p \lesssim 2.4$. The cutoff energy $E_{\rm{max}}$ depends on $z$, and is 
calculated according to Eq. (\ref{eq:max_energy}). The function $f(z)$ describes the injection profile along
the jet that defines the length of the acceleration region.

In most studies in the literature, the form of the injection function $Q (E)$ --- especially the spectral
index $p$ --- is left as a free parameter, whose value may be used as an indication of the dominant 
particle acceleration mechanism. In the case of $p \sim 2$, non-relativistic shocks present themselves
as a standard solution to achieve such a particle spectrum, while very hard spectra (down to $p \sim 1$) 
have been shown to be possible in the case of magnetic reconnection \citep[e.g.,][]{SironiSpitkovsky2014,Guoetal2016}. 
Diffusive shock acceleration at relativistic parallel shocks produces particle spectra with $\alpha \sim 2.2$ -- $2.3$
\citep[e.g.,][]{KirkDuffy1999}. Mildly relativistic, oblique shocks are capable of producing spectral indices 
in a wide range, from $p \sim 1$ to very steep spectra, depending primarily on the obliquity, i.e., the 
angle of the ordered component of the magnetic field with respect to the shock normal \citep{SummerlinBaring2012}. 
Thus, unfortunately, the diagnostic of determining $p$ from model fitting of spectral energy distributions 
of jet sources, is not sufficient to pin down the dominant particle acceleration mechanism, one of the fundamental
unknowns in our understanding of relativistic jets (but see some
additional discussion in Section~\ref{discussion}).

Finally, the value of the normalization constant $Q_0$ is obtained from the total power injected in 
each particle species

\begin{equation}
L_i = \int_V d\vec{r}\int_{E_{\rm{min}}}^{E_{\rm{max}}}\,dE\, E\,Q_i(E,z),
\label{eq:calculation_q0}
\end{equation}

\noindent where the index $i$ runs over protons and electrons. 
The interaction of relativistic protons with matter and radiation injects charged pions, muons, and secondary 
electron-positron pairs. Electron-positron pairs are also created through photon-photon annihilation. These 
processes define the specific expressions for $Q(E,z)$ for the respective secondary particles. 

The sources of charged pions are proton-photon and proton-proton interactions. Suitable expressions for the 
corresponding pion injection functions were presented in  \cite{AtoyanDermer2003}, \cite{Kelneretal2006}, and 
\cite{KelnerAharonian2008}, while expressions for the consequent muon injection may be found in \cite{Liparietal2007}. 
Finally, electrons and positrons are injected through muon decay, photopion production, and $\gamma\gamma$
absorption. Convenient expressions for the electron/positron injection rate from muon decay are presented in
\cite{Schlickeiser2002}. Expressions for the pair injection rate from photopair production may be found in 
\cite{Chodorowskietal1992} and \cite{Muckeetal2000}. The rate of pair production from photon-photon 
annihilation

\begin{equation}
\gamma+\gamma\rightarrow e^++e^-.
	\label{eq:gamma-gamma}
\end{equation}

\noindent is derived in \cite{BoettcherSchlickeiser1997}. Two-photon annihilation is also a photon sink. Its 
effect on the escape of photons from the jet is discussed below (Section \ref{sec:radiation}).

The evolution of the particle distributions is governed by transport equations of the form

$$
\frac{\partial N(E,t,z)}{\partial t} 
+ \frac{\partial\left( \Gamma_{\rm jet} v_{\rm jet} N(E,t,z)\right)}{\partial z} + 
\frac{\partial \left( b(E,t,z) N(E,t,z) \right) }{\partial E} 
$$
\begin{equation}
- {\partial \over \partial E} \left[ {E^2 \over (a + 2) \, t_{\rm acc}} \, {\partial N (E, t, z) \over 
\partial E} \right] + \frac{N(E,t,z)}{ T_{\rm dec}(E) } + \frac{N(E,t,z)}{T_{\rm esc}(E) } = Q(E,t,z)
\label{eq:transport_equation_t}
\end{equation}

The second term on the l.h.s. of Eq. (\ref{eq:transport_equation_t}) represents particle convection along
the jet, the third term describes systematic energy losses/gains, where $b(E,z)=-dE/dt|_{\rm tot}$ 
encompasses all energy loss plus first-order acceleration terms. The fourth term on the l.h.s. 
represents diffusion in momentum space and describes second-order (stochastic) shock acceleration
on a characteristic time scale $t_{\rm acc}$. Here, $a = v_s^2 / v_A^2$ is the square of the ratio 
of shock speed to the Alfv\'en speed in the co-moving jet plasma. $T_{\rm dec}$ is the decay time
scale. This term vanishes for stable particles. However, a corresponding term should be included
for electrons and positrons to account for the loss of pairs due to pair annihilation. $T_{\rm esc}$ 
is the particle escape time scale. 

This equation is appropriate for a very general case of a spatially extended, inhomogeneous jet,
in which the physical parameters may vary in time and space along the jet. In time-independent 
(steady-state) models, the first term $\partial N / \partial t$ is set to zero. Homogeneous, 
single-zone models neglect the second (convection) term. In many treatments, momentum diffusion
(and, thus, stochastic acceleration) is also neglected, thus leaving out the fourth term on the 
l.h.s. Note that Eq. (\ref{eq:transport_equation_t}) still assumes that the jet is homogeneous 
across the jet cross section (i.e., no radial dependence) and that the particle distributions 
are locally isotropic in the co-moving jet frame. Also, spatial diffusion has been neglected
\citep[see, e.g.,][for a blazar jet model that includes spatial diffusion]{Chenetal2015,Chenetal2016}.

\subsection{Radiative output}
\label{sec:radiation}

We now proceed to calculate the total radiative spectrum of all particle species produced by each of 
the interaction processes described in Sect. \ref{sec:microphysics}. In this exposition, we will 
restrict ourselves to quoting simple, $\delta$-function approximations for the respective emissivities.
Detailed expressions for the emissivities can be found, e.g., in 
\cite{AtoyanDermer2003,RomeroVila2008,ReynosoRomero2009,DermerMenon2009,Boettcheretal2012}, and 
references therein. 

The emissivities $j (\epsilon) = dE/(dV \, dt \, d\epsilon \, d\Omega)$ [erg cm$^{-3}$ s$^{-1}$] at 
photon energy $h \nu = \epsilon \, m_e c^2$ are commonly first evaluated in the co-moving jet frame, 
and then transformed to the observer frame. The relevant transformations are governed by the Doppler 
factor

\begin{equation}
D=\left[\Gamma_{\rm{jet}}\left(1-\beta_{\rm{jet}}\cos\theta_{\rm{jet}}\right)\right]^{-1}
	\label{eq:boost-factor}
\end{equation}
\noindent where $\theta_{\rm{jet}}$ is the jet viewing angle and $\beta_{\rm{jet}}=v_{\rm{jet}}/c$.

Relativistic Doppler effects then result in photon energies being blue/red shifted as $\epsilon_{\rm obs}
= D \, \epsilon$, and variability time scales appear contracted as $\delta t_{\rm obs} = \delta t / D$.  
The relativistic boosting of radiative flux depends on the geometry \citep[see, e.g.,][]{BlandfordLind1985}.  
In the case of a continuous, extended jet, the flux will be enhanced by a factor of $D^2$ so that the 
observed flux can be calculated as 

\begin{equation}
	F_{\epsilon_{\rm obs}, \rm ext} = {1 \over d_L^2} \int_{z_{\rm{acc}}}^{z_{\rm{max}}} r_{\rm{jet}}^2(z) 
	D^2 (z) \, j (\epsilon_{\rm obs}/D, z) \, dz.
	\label{eq:luminosity}
\end{equation}
\noindent where $d_L$ is the luminosity distance to the source.

In the case of single emission zone with co-moving volume $V$, the flux transformation is determined by
a factor $D^3$, i.e.,

\begin{equation}
F_{\epsilon_{\rm obs}, \rm zone} = {V \, j (\epsilon_{\rm obs} / D) \, D^3 \over d_L^2}
\label{eq:flux_singlezone}
\end{equation}

\noindent where the additional factor of $D$ originates from the volume integration (which is taken over 
$z$ in the laboratory frame in Eq. \ref{eq:luminosity}), taking into account Lorentz contraction and 
light-travel-time effects. 

The dominant radiation mechanisms resulting from high-energy,
accelerated particles are: synchrotron radiation (of primary electrons and protons as well as
secondary pions, muons, and electrons/positrons), inverse Compton scattering (relevant only for electrons
and po\-si\-trons), and $\pi^0$ decay. 

The synchrotron spectrum from an ensemble of charged particle of mass $m$, charge $q = Z \, e$, and energy 
$\gamma \, m c^2$ moving at random directions with respect to a magnetic field $B = B_{\rm G}$~G peaks at 
a frequency

\begin{equation}
\nu_c^{\rm sy} = {3 q B \over 4 \pi m c} \, \gamma^2 \equiv \nu_0 \, \gamma^2 
\approx 4.2 \times 10^6 \, B_{\rm G} \gamma^2 \, Z \, {m_e \over m} \; {\rm Hz}
\label{eq:nu_sy}
\end{equation}

Based on a simple $\delta$-function approximation for the synchrotron emissivity of a single particle, the
synchrotron spectrum of an ensemble of particles with energy distribution $n (\gamma)$ can be evaluated 
as\footnote{note a typo in Eqs. (3.38) and (3.39) in \cite{Boettcheretal2012}, where a factor $c$ is missing}

\begin{equation}
j_{\nu}^{{\rm sy}, \delta} = {4 \over 9} \, c \, \left( {q^2 \over m \, c^2} \right)^2 \, U_{\rm mag} \, 
\nu_0^{-3/2} \, \nu^{1/2} \, n\left( \sqrt{\nu \over \nu_0} \right).
\label{eq:jsy}
\end{equation} 

In the Thomson regime ($\epsilon \, \gamma \ll 1$), a target photon of energy $\epsilon$ will be scattered 
to higher energies by an isotropic ensemble of relativistic electrons of energy $\gamma \, m_e c^2$ to an 
average scattered photon energy $\epsilon_s \approx \gamma^2 \, \epsilon$. 
The Compton emissivity resulting 
from Compton scattering of a target photon field with photon number distribution $n_{\rm ph} (\epsilon)$ by 
a relativistic electron distribution $n_e (\gamma)$ in the Thomson regime can be approximated as

\begin{equation}
j_{\nu}^{{\rm Thomson}, \delta} (\epsilon_s) \approx {h \, c \, \sigma_T \, \epsilon_s^2 \over 8 \pi}
\, \int\limits_{\epsilon_s}^{\infty} d\gamma \, {n_e (\gamma) \over \gamma^4} \, \int\limits_{\epsilon_s
/ (2 \gamma^2)}^{\infty} d\epsilon \, {n_{\rm ph} (\epsilon) \over \epsilon^2}
\label{eq:jCompton}
\end{equation}

where a simple $\delta$ function approximation for the Compton cross section has been employed.
More general expressions for the synchrotron and Compton emissivities, including Compton scattering in the
Klein-Nishina regime, may be found, e.g., in \cite{BlumenthalGould1970,RybickiLightman1979,DermerMenon2009,Boettcheretal2012}.

In the decay of a $\pi^0$ meson of energy $E_{\pi^0} = \gamma_{\pi^0} m_{\pi} c^2$, two photons with energy
$E_{\gamma} \approx (1/2) E_{\pi^0}$ are produced, while the production of $\pi^0$ mesons depends on the 
(energy-dependent) $p\gamma$ cross section, branching ratio (i.e., fraction of $p\gamma$ pion production 
events resulting in the production of a $\pi^0$) and inelasticity (i.e., the fraction of the proton's energy
transferred to the newly produced pion). A convenient approximation for the calculation of the resulting 
photon production spectrum can be found in \cite{AtoyanDermer2003,RomeroVila2008}.

The photon emission spectrum must be corrected for various absorption processes. At low (radio) frequencies, 
synchrotron-self absorption (SSA) will be dominant \citep[see, e.g.,][]{RybickiLightman1979,Boettcheretal2012}, 
while at high energies ($E \gg 1$~GeV) $\gamma$-ray absorption can be caused by photon-photon annihilation into 
electron-positron pairs (see Eq. (\ref{eq:gamma-gamma})). The $\gamma\gamma$ opacity for a photon of energy 
$\epsilon_{\gamma}$ in a target photon field $n_{\rm ph} (\epsilon)$ is

\begin{equation}
\kappa_{\gamma\gamma}(\epsilon_{\gamma}) = \frac{1}{2} \int\limits^{1}_{-1} d\mu 
\int\limits^{\infty}_{\epsilon_{\rm th}} d\epsilon \, \sigma_{\gamma\gamma}(\epsilon_{\gamma}, \epsilon, \mu)
\, n_{\rm ph} (\epsilon) \, (1 - \mu) 
\label{eq:kappa_gamma-gamma}
\end{equation}

\noindent where $\mu$ is the cosine of the collision angle, and $\sigma_{\gamma\gamma}$ is the annihilation 
cross-section \citep[e.g.,][]{GouldSchreder1967}, and $\epsilon_{\rm th}$ is the threshold energy,

\begin{equation}
\epsilon_{\rm{th}} = \frac{2}{\epsilon_\gamma \, (1 - \mu)}. 
\label{eq:gamma-gamma-threshold}
\end{equation} 

The $\gamma\gamma$ optical depth $\tau_{\gamma\gamma}$ is then evaluated as the length integral over 
$\kappa_{\gamma\gamma}$ along the photon propagation direction. Eq. (\ref{eq:kappa_gamma-gamma}) is 
simplified when the target photon distribution is isotropic. A convenient expression for 
$\tau_{\gamma\gamma}$ in this particular case is obtained in \cite{GouldSchreder1967}.

If the absorbing photon field is located outside the emission region (as, e.g., in the case of VHE $\gamma$-ray
absorption by the Extragalactic Background Light, see Section \ref{blazars}), the $\gamma\gamma$ opacity simply 
leads to a correction factor $exp(-\tau_{\gamma\gamma})$ with respect to the unabsorbed flux spectrum, 
$F_{\rm intr} (\epsilon)$. However, in case emission and absorption are taking place co-spatially, the 
solution to the radiative transfer equation becomes more involved. In the case of a homogeneous source, 
this solution is given by

\begin{equation}
F_{\rm obs}^{\rm int} (\epsilon_{\gamma}) = F_{\rm intr} (\epsilon_{\gamma}) \, {1 - e^{-\tau_{\gamma\gamma}
(\epsilon_{\gamma})} \over \tau_{\gamma\gamma} (\epsilon_{\gamma})}
\label{eq:Fabs_internal}
\end{equation}

In the optically-thin limit ($\tau_{\gamma\gamma} \ll 1$), Eq. (\ref{eq:Fabs_internal}) simply reduces to
$F_{\rm obs}^{\rm opt. \, thin} = F_{\rm intr}$, as expected, while in the optically-thick limit 
($\tau_{\gamma\gamma} \gg 1$), it becomes $F_{\rm obs}^{\rm opt. \, thick} = F_{\rm intr} / \tau_{\gamma\gamma}$.
If the target photon field has a power-law spectrum $n_{\rm ph} (\epsilon) \propto \epsilon^{-(\alpha_t + 1)}$
(i.e., an energy index of $\alpha_t$), then the opacity also has a power-law depencence on energy, 
$\tau_{\gamma\gamma} (\epsilon_{\gamma}) \propto \epsilon_{\gamma}^{\alpha_t}$. Consequently, internal
$\gamma\gamma$ absorption leads to a spectral break by $\Delta\alpha_{\gamma} = \alpha_t$ around the 
photon energy where $\tau_{\gamma\gamma} (\epsilon_{\gamma}) = 1$. The presence or absence of such
spectral breaks in the $\gamma$-ray spectra of relativistic jet sources can be used to constrain the
bulk Lorentz factor of the emission region \citep[e.g.,][]{Baring1993,DondiGhisellini1995,Boettcheretal2012}.
As we will elaborate on in Section \ref{location}, $\gamma\gamma$ absorption in the immediate environment
(especially, the Broad Line Region) of powerful blazars also provides constraints on the location of the 
$\gamma$-ray emission region in those objects. 

In addition to photons, if hadronic charged-pion production processes are relevant, also electron and
muon neutrinos at $\sim$~TeV -- PeV energies are produced. The detection of astrophysical neutrinos
and their unique identification with relativistic jet sources would therefore uniquely identify sites
where protons are being accelerated to ultrarelavisitic energies exceeding the pion production threshold
(see Eq. \ref{pionthreshold}), possibly even reaching the energies of ultra-high-energy cosmic rays 
(UHECRs) with energies $E \gtrsim 10^{19}$~eV. 

Convenient templates for the pion production spectra from relativistic $p\gamma$ and $p-p$ interactions 
in the case where pion and muon synchrotron losses may be neglected, can be found in 
\cite{KelnerAharonian2008,Kelneretal2006}, respectively. If pion and muon synchrotron emission can
not be neglected, their energy losses prior to decay need to be taken into account. A formalism to
evaluate the secondary-particle production rates in this more general case, has been developed in
\cite{Huemmeretal2010}.

\subsection{Polarization}
\label{sec:polarization}

Relativistic jets are expected to produce polarized synchrotron radiation. Such radiation is observed from 
radio to X-rays. At higher energies, inverse Compton upscattering of the polarized synchrotron photons can 
yield polarized X/gamma-rays \citep[e.g.,][]{BonomettoSaggion1973}, and also anisotropic Compton scattering
of unpolarized radiation fields by non-re\-la\-ti\-vistic or mildly relativistic electrons can introduce 
polarization \citep[e.g.,][]{Poutanen1994,Krawczynski2012}. Evidence for polarization in high-energy radiation 
from relativistic jet sources has been found, e.g., in Cygnus X-1 \citep{Laurentetal2011,Romeroetal2014}, and 
there are several developments for X-ray and $\gamma$-ray polarimeters currently on-going and proposed, such 
as the X-Ray Imaging Polarimeter Experiment \citep[XIPE, ][]{Soffitta2013}, or the Polarization Spectroscopic 
Telescope Array \citep[PolSTAR, ][]{Krawczynski2016}. 

Both in the case of optically-thin synchrotron and synchrotron-self-Compton scattering, the preferred 
polarization direction indicates the predominant orientation of the magnetic field, as the Electric Vector 
Position Angle (EVPA) is perpendicular to the projection of the B-field onto the plane of the sky. A 
non-vanishing synchrotron polarization therefore clearly indicates an at least partially ordered magnetic 
field. If the B-field is perfectly ordered, then optically-thin synchrotron radiation by a non-thermal 
power-law particle distribution with index $p$ yields a degree of polarization $\Pi$ of

\begin{equation}
\Pi_{\rm Sy} = {p + 1 \over p + {7 \over 3}} 
\label{eq:Pi_synchrotron}
\end{equation}

\noindent corresponding to polarization degrees of $\Pi \sim 70$ -- 75~\% for typical spectral indices of
$p \sim 2$ -- 3 \citep[e.g.,][]{RybickiLightman1979}. The measured degree of optical polarization in relativistic 
jet sources is typically in the range $\Pi \lesssim 30$~\%, and often significantly smaller than that. This 
indicates that the emission must originate from regions with partially disordered magnetic fields. 

Especially in the case of blazars, it is known that the polarization properties are highly variable, both in
degree of polarization and polarization angle (PA). Paritcularly noteworthy in this context are large swings 
in the PA \citep{Romeroetal1995}, which sometimes correlate with multi-wavelength flares 
\citep{Marscheretal2008,Marscheretal2010,Abdoetal2010,Blinovetal2015,Blinovetal2016}. Such PA swings naturally 
suggest the existence of helical fields \citep[e.g.,][]{Zhangetal2014,Zhangetal2016}. Models based on random 
fluctuations in a turbulent field have been also proposed \citep{Marscher2014}.

Time-dependent multiwavelength polarization measurements diagnose the topology of the magnetic field 
in the emission region and its change during variability events. Such observations may therefore be a
unique tool to identify the role of magnetic fields in the formation of jets and in the relativistic 
particle acceleration processes leading to the emission of high-energy radiation. The diagnostic power 
of (time-dependent) polarization signatures in blazars will be discussed in more detail in Section 
\ref{blazars}.

At low (radio) frequencies, the propagation of polarized emission through a magnetized medium leads
to a rotation of the EVPA, known as Faraday Rotation \citep[e.g.][]{Pacholczyk1963}. This may occur even 
within the source, and thus lead to effective Faraday depolarization \citep[e.g.,][]{Burn1966}, as the 
EVPAs of radiation produced in different locations along the line of sight within the source appear 
rotated by different amounts. The amount of Faraday rotation, $\Delta\psi$, i.e., the angle by which 
the EVPA is rotated, depends on the electron density $n_e$, the projection of the magnetic field 
$\overrightarrow{B}$ along the line of sight $\overrightarrow{s}$, and the radiation wavelength 
$\lambda$ as 

\begin{equation}
\Delta\psi \propto \lambda^2 \, \int n_e(s) \overrightarrow{B}(s) \cdot d\overrightarrow{s}
\label{Faraday}
\end{equation}
Due to the $\lambda^2$ scaling, this effect is generally only relevant at radio frequencies and is
negligible at optical or higher frequencies.

\section{Blazars}

\label{blazars}

The class of peculiar extragalactic sources dubbed {\it blazars} represents a quite small, but remarkably 
interesting fraction of the entire population of AGN (e.g. \citealt{UrryPadovani1995}). 
While a formal definition is difficult to establish, the canonical features used in their classification 
include the presence of a compact radio core, with flat or even inverted spectrum, extreme variability 
(both in timescale and in amplitude) at all frequencies, and a high degree of optical and radio polarization. 
More recently, intense gamma-ray emission (although not ubiquitous, but in many cases dominating the total
bolometric radiative power output) has also been used for blazar selection. A recent census counts about 
3000 known blazars \citep{Massaroetal2009}.

Phenomenologically, blazars are further divided in two subgroups depending on special observational
characteristics: 1) {\it BL Lacertae} objects are characterized by an extreme weakness (equivalent width 
$<5$ \AA) or even absence of emission lines in their optical spectra, a feature generally attributed to a 
low accretion rate onto the central supermassive black hole and an environment largely depleted of potential
fuel for black-hole accretion, not exhibiting a prominent broad-line region; 2) if the optical spectra display 
broad emission lines typical of quasars (or QSOs = Quasi Stellar Objects), blazars will be instead classified 
as {\it Flat Spectrum Radio Quasars} (FSRQ). FSRQs are generally more powerful than BL Lacs. The FSRQ population 
follows a positive cosmological evolution close to that of normal quasars, being quite rare in the local Universe 
and with a density peaking around $z\sim2$ \citep[e.g.,][]{DunlopPeacock1990}. The BL Lac population, instead, 
displays a quite different trend. In particular, the less powerful high-frequency peaked BL Lacs (HBL), follow 
a {\it negative} cosmological evolution \citep{Rector2000,Ajelloetal2014}, i.e. they tend to be more numerous 
in the local Universe, with their density decreasing with redshift. The different evolutions could be related 
to a generic connection between FSRQs and BL Lacs, the latter representing the final stage of the FSRQ activity, 
after the fuel for black-hole accretion has been exhausted \citep{Cavaliere2002,BoettcherDermer2002}. 

Relativistic beaming offers a simple explanation for the peculiarities displayed by blazars 
\citep{Blandfordrees1978,BlandfordKoenigl1979}, in particular the inverted radio spectrum, the huge brightness 
temperature, the existence of superluminal moving features, the short variability timescale and the powerful 
$\gamma$-ray emission. Different methods agree to determine typical Doppler factors of localized emission regions
in blazar jets in the range  $D\simeq 10-20$ (but reaching in some cases values as large as 50), implying a 
similar bulk Lorentz factor for the plasma outflowing in the jet 
\citep[][and references therin]{Listeretal2013,GhiselliniTavecchio2015} and a viewing angle as small as few 
degrees. Blazars, therefore, represent radio sources in which the jet velocity points almost along the line of 
sight. In this geometry, the relativistically beamed non-thermal continuum produced within the jet easily 
outshines any other emission component from the active nucleus or the host galaxy. In the view of the 
{\it unification scheme}, BL Lacs and FSRQ are considered the beamed counterparts of the FRI (low power) 
and FRII (high power) radiogalaxies, respectively \citep[e.g.,][]{UrryPadovani1995}.

\begin{figure}[!htp]
\hspace{1.5truecm}
\includegraphics[trim = 0mm 5mm 0mm 0mm, clip,width=9cm,angle=0]{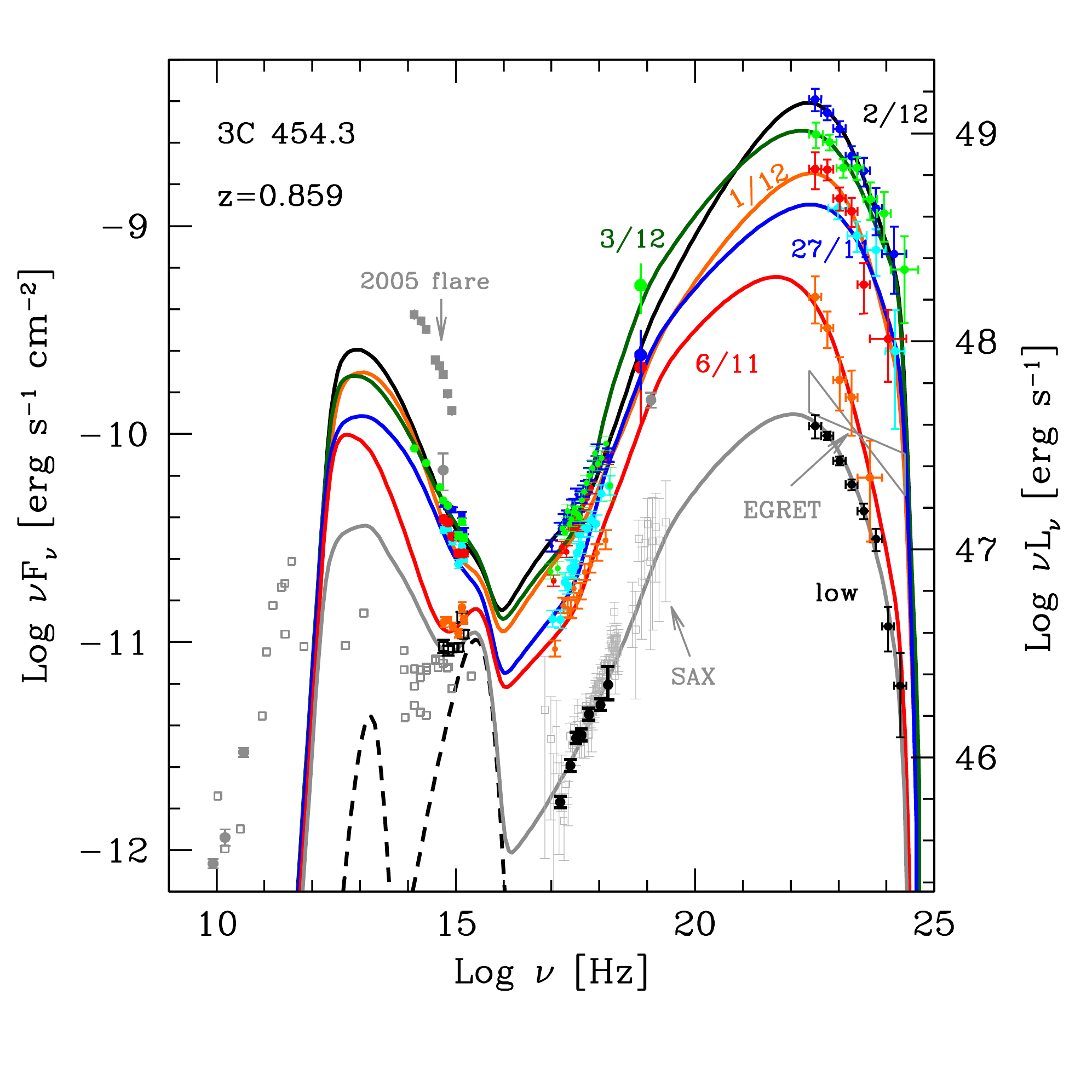}
\caption{Spectral energy distributions of the FSRQ 3C454.3 during different activity states in 2012 (coloured) 
and a quiescent phase (gray). The data clearly show the typical ``double humped" shape of the SED. Curves show 
the result of a one-zone leptonic emission model. The reader can appreciate the extreme variability, especially 
in the $\gamma$-ray band. From \cite{Bonnolietal2011}.}
\label{FigSED3C454}
\end{figure}

The remarkably smooth continuum emission of blazars, extending over the entire electromagnetic spectrum, from 
the radio band to $\gamma$-ray energies (see Fig.\ref{FigSED3C454} for an example), is characterized in the 
$\nu F(\nu )$ representation (the so-called spectral energy distribution, SED) by a typical ``double humped'' 
shape \citep[e.g.,][]{Sambrunaetal1996,Fossatietal1998}. While the characteristics of the low energy component 
were already well known and fairly well understood as synchrotron radiation of relativistic electrons in the
1980s \citep[e.g.,][]{MarscherGear1985}, it was realized only after the advent of the $\gamma$-ray telescope 
EGRET (Energetic Gamma-Ray Experiment Telescope) onboard the {\it Compton Gamma-ray Observatory} in the 1990s 
\citep{Fichteletal1994} that many blazars exhibit significant $\gamma$-ray emission, often dominating the SEDs. 
Including the $\gamma$-ray output, the apparent bolometric luminosities of blazars reach, in some 
cases, $L_{\gamma}=10^{49}$ erg s$^{-1}$ \citep[e.g.,][]{Bonnolietal2011}. These extreme luminosities, combined
with the observed short (often intra-day) variability, provide one line of evidence for relativistic beaming in
blazars \citep[see, e.g.,][for a review]{Schlickeiser1996}.

Improved $\gamma$-ray instruments in space ({\it Fermi} and {\it AGILE}) and the development of Cherenkov 
telescope facilities on the ground (H.E.S.S., MAGIC, VERITAS, and the future CTA), together with intense 
multifrequency coordination efforts provide increasingly complete coverage of the rapidly changing SEDs,  
which is instrumental in the understanding of the physical processes shaping the blazar phenomenon. The 
Cherenkov telescopes, sensitive to very high energy (VHE) $\gamma$-rays ($E>50$ GeV), have revealed an 
important population of blazars (BL Lacs, in particular) characterized by an intense emission at TeV 
energies which, besides jet physics, can be exploited as a probe for cosmic backgrounds and intergalactic 
magnetic field.

\subsection{Emission models: the standard scenario}
\label{standardscenario}

As clearly indicated by the high degree of radio and optical polarization, the low-energy
spectral component of blazars is produced through the synchrotron mechanism, most likely by a
population of highly relativistic electrons. The origin of the high-energy component is still 
debated. Two general scenarios for the production of high energy radiation are discussed 
(see also Sec.~\ref{model}), namely {\it hadronic} or {\it leptonic} models 
\citep[e.g.,][]{Boettcheretal2013}.

\subsubsection{Leptonic models}
\label{leptonic}

The most popular class of models adopts the leptonic scenario, in which
the high-energy component is the result of the inverse Compton
scattering of soft photons by the same electrons responsible for the
synchrotron emission. Different sources of soft photons can be
considered; the emission will be dominated by inverse Compton scattering of
the photon component with the largest energy density (as measured in
the co-moving jet frame). Depending on the nature of soft photons
dominating the IC process, the leptonic high-energy emission is
comprised of SSC or {\it External Compton} (EC) radiation.

In the SSC model \citep[e.g.,][]{Maraschietal1992,Tavecchioetal1998} it is assumed that 
target photons are dominated by the synchrotron photons themselves. In this class of models 
synchrotron and IC emission are strongly coupled. This affords the possibility to derive 
robust constraints on the basic physical quantities of the jet from the observed SED 
\citep{Tavecchioetal1998}. In the EC model one assumes that the low energy radiation produced 
in the central regions of the AGN (directly emitted by the accretion disk or reprocessed by 
the gas in the Broad Line Region or by the parsec-scale molecular torus) dominates over the 
synchrotron radiation. However, even in this case, SSC emission is still expected to make a
non-negligible contribution, especially at X-ray energies. 
A possibility \citep{DermerSchlickeiser1993} is thus to consider the direct UV emission from the
accretion disk; however for sufficiently large distances from the disk, the de-beaming suffered by
the radiation directly coming from the disk causes a strong depression of this contribution in the 
jet frame. The primary disk radiation is partly reflected/reprocessed by the gas of the BLR and 
beamed in the jet frame, providing a strong contribution to the emission \citep{Sikoraetal1994}.
\cite{Blazejovskietal2000} pointed out that IC scattering of the thermal near-infrared radiation 
($T\simeq 10^3$ K) emitted by the dust of the parsec-scale torus expected to surround the central 
AGN region could provide the dominant contribution to the high energy emission, especially 
in the energy band 10 keV -- 100 MeV. This could be the case especially for FSRQ detected above few 
tens of GeV (see below), however for BL Lacs this component is less likely to be
important as recent evidence suggests the dusty torus dwindles at lower
accretion rates \citep{Plotkinetal2012b}.  

Not only the optical/UV emission from the accretion disk, but also the beamed synchrotron emission 
from the jet can be ``reflected'' by the BLR and thus serve as target photon field for Compton
scattering. The double change of frame (source $\rightarrow $ lab frame $\rightarrow $ source) 
leads to a great amplification of the energy density of the soft radiation 
\citep{GhiselliniMadau1996,Tavanietal2015}. However, as discussed by \cite{BoettcherDermer1998} 
and \cite{Bednarek1998}, when the constraints posed by the travel time of the radiation are taken 
into account, the relative contribution of this mechanism to the total emission may be severely 
suppressed. \\

The question of the dominant target photon field is intimately linked to the location of the emission 
region along the jet, which will provide additional clues towards the dominant particle acceleration
mechanism. This aspect will be discussed in more detail in Section \ref{location}

\begin{figure}[!htp]
\includegraphics[trim = 0mm 5mm 0mm 30mm, clip,width=11cm,angle=0]{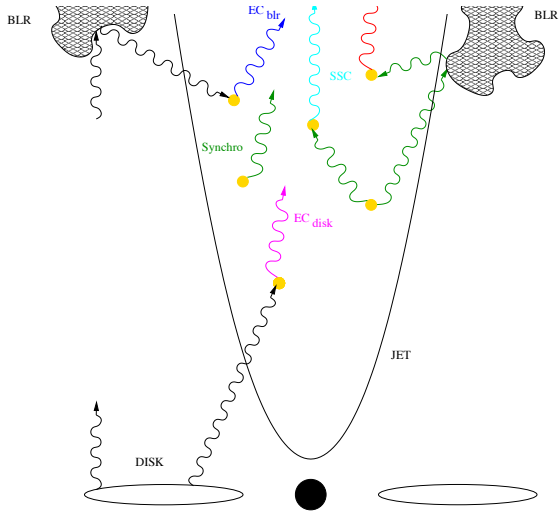}
\caption{Sketch illustrating the basic emission model for blazars. Relativistic electrons in the jet 
produce radiation through synchrotron emission and IC scattering. Soft target photons for the IC process 
are the synchrotron photons themselves ({\it SSC radiation}) and those produced in the external environment 
({\it EC radiation}), namely produced in the disk, or reprocessed in the BLR. A third class of models 
({\it the mirror model}) consider also the synchrotron photons ``reflected'' by the BLR. Further out 
also the IR radiation emitted by the molecular torus can be scattered.}
\label{FigBlazarModels}
\end{figure}

The general set-up of leptonic blazar emission models is sketched in Fig.\ref{FigBlazarModels}. The jet 
flow is expected to become dissipative around $10^2$ -- $10^3$ Schwarzschild radii (corresponding to about 
0.01 -- 0.1 parsec for a BH mass of $10^9$ M$_{\odot}$), likely through shocks or magnetic reconnection 
events which can accelerate particles to ultrarelativistic energies (sometimes this is defined as the 
``blazar zone"). Typical variability timescales of $\delta t_{\rm var} \sim $~few hours (although shorter 
timescales events have been occasionally observed), together with typical Doppler factors of $D \sim 10$ -- 20, 
limit the size of the emission region to $R \lesssim \delta t_{\rm var} \, c \, D / (1 + z) \sim 
10^{15}$ -- $10^{16}$~cm.

\begin{figure}[!htp]
\hspace{1.truecm}
\includegraphics[trim = 0mm 5mm 0mm 0mm, clip,width=9cm,angle=0]{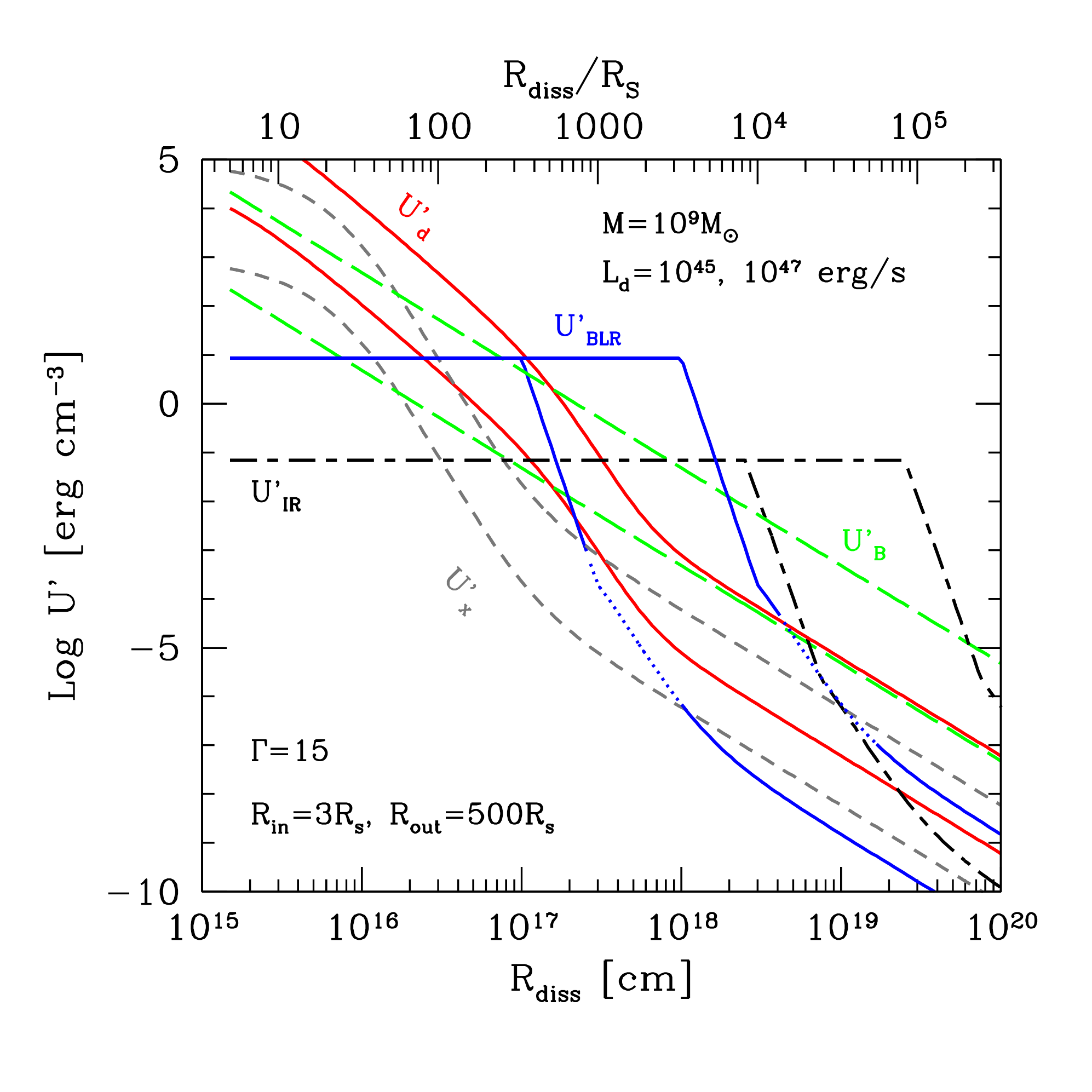}
\caption{Comparison of different energy densities  as measured in the jet reference frame.
The jet  is assumed to have a bulk Lorentz factor $\Gamma =15$ and the black hole mass is assumed 
to be $M=10^9 M_\odot$.  The two sets of lines correspond to two disk luminosities: 
$L_{\rm d}=10^{45}$ and $10^{47}$ erg s$^{-1}$. The magnetic energy density (long dashed lines) 
is calculated assuming that a fraction $\epsilon=0.1$ of the jet power is carried as magnetic 
field. From \cite{GhiselliniTavecchio2009}. 
  }
\label{FigUrad}
\end{figure}

SED modeling of a large number of blazars \citep[e.g.,][]{Ghisellinietal2010} suggests that 
SSC and EC processes dominate the high energy emission of different types of blazars: in 
the case of FSRQ, which display bright thermal features (emission lines and the {\it blue bump}), 
$\gamma$-ray production is likely dominated by EC (often with a contribution from SSC in the X-ray 
band), while in high-frequency peaked BL Lac objects (characterized by very weak or even absent 
thermal components) the SSC process dominates. However, the situation is less straightforward 
for intermediate (relatively powerful) BL Lac objects, in which the EC emission could play 
some role \citep{Boettcher2010}. 
Moreover, for FSRQ, the actual external radiation field ruling the EC emission depends on the position 
of the emitting region along the jet. This is shown in Fig. \ref{FigUrad}, in which the energy densities 
of the various radiation fields (in the rest frame of the jet) are plotted as a function of the distance 
from the central engine. Within the radius marking the distance where the BLR clouds are 
located\footnote{For simplicity we are assuming that the clouds producing the broad lines are 
located at the same distance. In reality, different emission lines are likely produced in a 
stratified BLR, in which the ionization factor decreases with distance \citep{Netzer2008}.} 
the energy density is dominated by the constant BLR radiation field. Beyond the BLR radius, 
$U_{\rm BLR}$ falls rapidly with distance and the radiation from the torus starts to dominate. 
The great majority of FSRQs can be modelled assuming that the emitting region is at distances 
smaller than the BLR radius \citep[e.g.,][]{Ghisellinietal2010,Ghisellinietal2014}, thus 
considering only EC with the BLR photons. However, for FSRQ detected at VHE, the emission 
likely occurs beyond this radius due to $\gamma\gamma$ absorption constraints (see Section
\ref{location}).

\subsubsection{Hadronic models}
\label{hadronic}

Even in the case of the leptonic models described above, protons are expected to be present in
the jet in order to ensure charge neutrality, 
unless a pure electron-positron pair plasma is assumed.
However, in order to contribute significantly to
the radiative output (through proton synchrotron and $p\gamma$ pion production), they need to be 
accelerated to energies $E_p \gg 10^{16}$~eV (see Section \ref{sec:microphysics}). In order for
such acceleration to be possible within the constrained size of the blazar emission region, large
magnetic fields of $B \gtrsim 10$~G are required. 

If the acceleration of protons to the required energies does happen in blazar jets, the hadronic
radiation processes discussed in Sections \ref{sec:microphysics} and \ref{sec:radiation} become
relevant. Due to the high magnetic fields, the radiative output from leptons in the emission region
will then be strongly dominated by synchrotron radiation, so that the high-energy (X-ray through
$\gamma$-ray emission is likely dominated by proton synchrotron radiation and synchrotron emission
of secondaries resulting from photopion production. Such models are termed {\it hadronic} or
{\it lepto-hadronic} models 
\citep[e.g.,][]{Mannheim1993,MannheimBiermann1992,MueckeProtheroe2001,Mueckeetal2003,Reynosoetal2011,PetropoulouMastichiadis2012,Boettcheretal2013,Petropoulouetal2015}

As the particle densities in blazar jets are very low
($n \lesssim 10^3$~cm$^{-3}$), pion production processes are
strongly dominated by p$\gamma$ interactions, in which case the
secondary particles (pions, muons, electrons/positrons) are produced
with large Lorentz factors of $\gamma \gtrsim 10^7$. $\pi^0$ decay
photons and synchrotron photons from electrons/positrons at these
energies emerges at $\gg 1$~TeV energies, where the emission region is
highly opaque to $\gamma\gamma$ absorption, producing additional
electron/positron pairs in a cascading process.  Due to the extreme
energies in protons and the relatively low radiative efficiency of
protons compared to leptons, as well as the increased bulk mass to
accelerate, hadronic jet models of blazars typically require several
orders of magnitude larger jet powers than leptonic models for the
same blazars. In some cases, the required jet powers even exceed the
Eddington luminosity of the central supermassive black hole, thus
requiring a revision of our current understanding of the formation of
relativistic jets and related energy considerations \citep[see, e.g.,
Section \ref{sec:energetics};][]{ZdziarskiBoettcher2015}.

\begin{figure}[!htp]
\includegraphics[width=10cm]{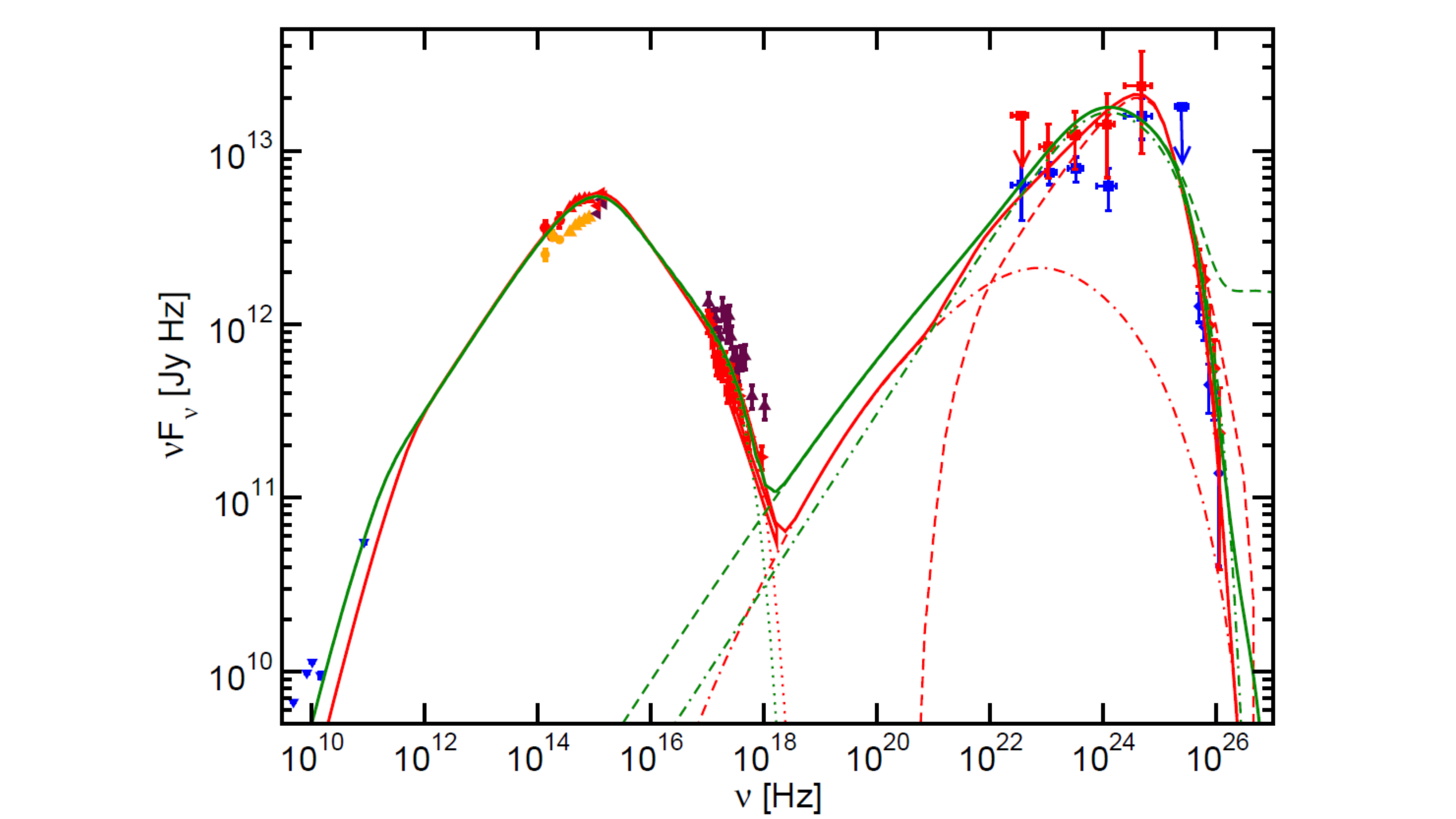}
\caption{Leptonic (red) and hadronic (green) fits to the simultaneous SED of the intermediate BL Lac 
object 3C66A. Different line styles indicate different radiation components: solid = total SED, including
correction for EBL absorption; dotted = primary electron synchrotron; red dashed = EC(torus); 
red dot-dashed = SSC; green dot-dashed = proton synchrotron; green dashed = total proton synchrotron +
cascade emission }
\label{3C66A_combined_fit}
\end{figure}

In terms of fitting SEDs of blazars, hadronic models have met with similar sucess as leptonic
models (see, e.g., Fig. \ref{3C66A_combined_fit}). Hence, SED modeling is generally not sufficient
to distinguish leptonic from hadronic models and, thus, to constrain the jet energetics, and additional
diagnostics need to be developed. One possibility lies in the different variability patterns expected
in leptonic and hadronic models, due to the vastly different radiative cooling times of protons
compared to electrons/positrons. Due to the complexity of hadronic models (including 8 different
particle species + photons, with a rather complicated energy dependence of p$\gamma$ pion
production processes), the development of time-dependent hadronic models has only recently
made significant progress \citep[e.g.,][]{Mastichiadisetal2013,Diltzetal2015,WeidingerSpanier2015}. 

Another alternative to distinguish leptonic from hadronic models consists of the fact that only
hadronic processes result in the production of high-energy neutrinos \citep[e.g.,][]{Petropoulouetal2015}. 
Thus, the identification of blazars with the sources of astrophysical TeV -- PeV neutrinos, as
now confidently detected by IceCube \citep{Aartsen2014} would unamibguously prove the presence
of ultrarelativistic protons in blazar jets. However, unfortunately, the angular resolution of
IceCube (and other neutrino observatories) is currently too poor to
provide such a unique identification (but see, e.g., \citealt{Kraussetal2016}).

In light of current developments of X-ray and $\gamma$-ray observatories with polarimetry capabilities,
high-energy polarization may also be considered as a distinguishing diagnostic of leptonic and hadronic
models. In particular, \cite{ZhangBoettcher2013} have shown that the X-ray and $\gamma$-ray emission 
from blazars in the case of a hadronic origin is expected to be polarized with polarization degrees 
similar to those observed in the optical band (i.e., $\Pi \lesssim$~a few tens of percent). On the 
other hand, in leptonic models, the X-ray emission of most (non-HBL) blazars is expected to be dominated 
by SSC emission, which is expected to exhibit a degree of polarization of about half that of the target 
photon field, i.e., $\Pi_X \sim \Pi_{\rm opt}/2$. In the case of EC dominated $\gamma$-ray emission,
those $\gamma$-rays are expected to be essentially unpolarized, in stark contrast to the prediction 
of hadronic models. Thus, X-ray and $\gamma$-ray polarimetry of blazars may serve as a powerful
diagnostic to distinguish leptonic and hadronic emission models.

\subsection{One-zone {\it vs} multizone/structured jets}
\label{structuredjets}

In the previous section we discussed the emission scenarios without specifying in detail the nature 
and the geometry of the emission region. Unfortunately, our knowledge of the geometry and even the 
composition of jets is still rather poor. Therefore in the modeling of blazars even the specific
characteristics of the region responsible for the emission should be considered as a free parameters.

Blazar models can generally be divided in inhomogeneous (or jet) and homogeneous models. In the first class of models 
\citep{BlandfordKoenigl1979,Ghisellinietal1985} one assumes that the observed emission is 
produced in an extended volume along the jet and a particular geometry (e.g. conical, parabolical) 
for it (see also Section \ref{sec:energetics}). Moreover one needs to specify the evolution of the 
main physical quantities (magnetic field, particle density, particle energy) along the jet: this 
necessarily introduces a great number of free parameters.

Most of the recent discussions in the literature consider a single, homogeneous emission region. 
The most direct support to these {\it one-zone} models comes from  the correlated variability at 
different frequencies \citep{Ulrichetal1997}, although deviations from this behavior 
\citep[e.g. the so-called orphan flares,][]{Krawczynskietal2004} are sometimes observed. 
It is clear that this is an over-simplification of the actual geometry and physics involved 
in jets, but with the advantage to drastically reduce the unknown parameters of the model. 
This approach is valid to approximate the very compact region where high-energy radiation is produced. 
Consequently, such models do typically not reproduce the low-frequency radio emission from blazars,
which is known to emerge in larger-scale ($\gg$~pc) jet structures, as evidenced by the drastically
different radio variability time scales compared to optical through $\gamma$-ray variability, and
consistent with the fact that such one-zone model configurations are optically thick to synchrotron
self-absorption at radio frequencies. 

Emission models typically assume a-priori that the jet contains high energy particles. An evident 
problem to consider is the identification of the physical processes able to energize particles to
such high energies. Clearly, the problem of the nature of the emission region and that concerning the 
acceleration process(es) are coupled. A ``paradigm'' popular until recent years is the so-called 
{\it shock-in-jet} model \citep{MarscherGear1985,Spadaetal2001}. It is assumed that, due to instability 
in the flow, an internal shock can form in the jet. Particles are then accelerated at the shock front 
through diffusive shock acceleration (Fermi I) processes. This scenario is supported by the observations 
of traveling ``knots" at VLBI ($\approx$ pc) scale. In this scheme the emitting region can be identified 
with the region downstream the shock, where particles injected from the front radiatively cool and emit 
\citep{Kirketal1998,ChiabergeGhisellini1999,Sokolovetal2004,BoettcherDermer2010,JoshiBoettcher2011,Chenetal2011}).

However, several criticisms to this view have been advanced. In fact several 
observational and theoretical arguments seem to challenge the shock model, at 
least in its simplest form. Among them, the observation of episodes of ultra-rapid 
variability in $\gamma$ rays both in BL Lacs and FSRQ, with timescales down to few 
minutes (\citealt{Aharonianetal2007},  \citealt{Albertetal2007}, \citealt{Arlenetal2013}, 
\citealt{Aleksicetal2011}), implies, through the causality argument, very compact emission 
regions ($R< c t_{\rm var} D\approx 10^{14} [t_{\rm var}/1 \,{\rm min}] \; D/10$ cm) 
incompatible with the expected size of a shock. Also, detailed fits often reveal that 
a model with a unique region is not able to provide a good representation of the SED 
and more than one component is required (\citealt{Aleksicetal2015}, \citealt{Barresetal2014}, 
\citealt{Tavecchioetal2011}). From a more fundamental  perspective, it is expected that, 
at the typical distance of the ``blazar zone", the flow is still dominated by the Poynting 
(i.e. magnetic) flux \citep[e.g.,][]{Komissarovetal2007}. In these circumstances diffusive 
particle acceleration by shocks is expected to be quite inefficient 
\citep{KennelCoroniti1984,Sironietal2015}.

All these issues bring stimulated new proposals for the structure of the emission 
region. Rapid variability suggests that (at least occasionally) the radiation is 
produced in very compact regions embedded in the jet. This have been identified 
with turbulent cells \citep{NarayanPiran2012,Marscher2014} or with 
large plasmoids resulting from efficient magnetic reconnection in a relativistic 
regime \citep{Gianniosetal2009,Giannios2013,Sironietal2015}. 
In both cases, the compact regions are thought to move relativistically in the jet 
reference frame. The resulting beaming of the emitted radiation in the observer 
frame (which is the combination of the beaming factor of the compact region in the 
jet frame and that of the jet flow with respect to the observer) can be very large, 
with Doppler factors easily reaching $D\sim50$. 

\begin{figure}[!htp]
\begin{center}
\includegraphics[width=9cm]{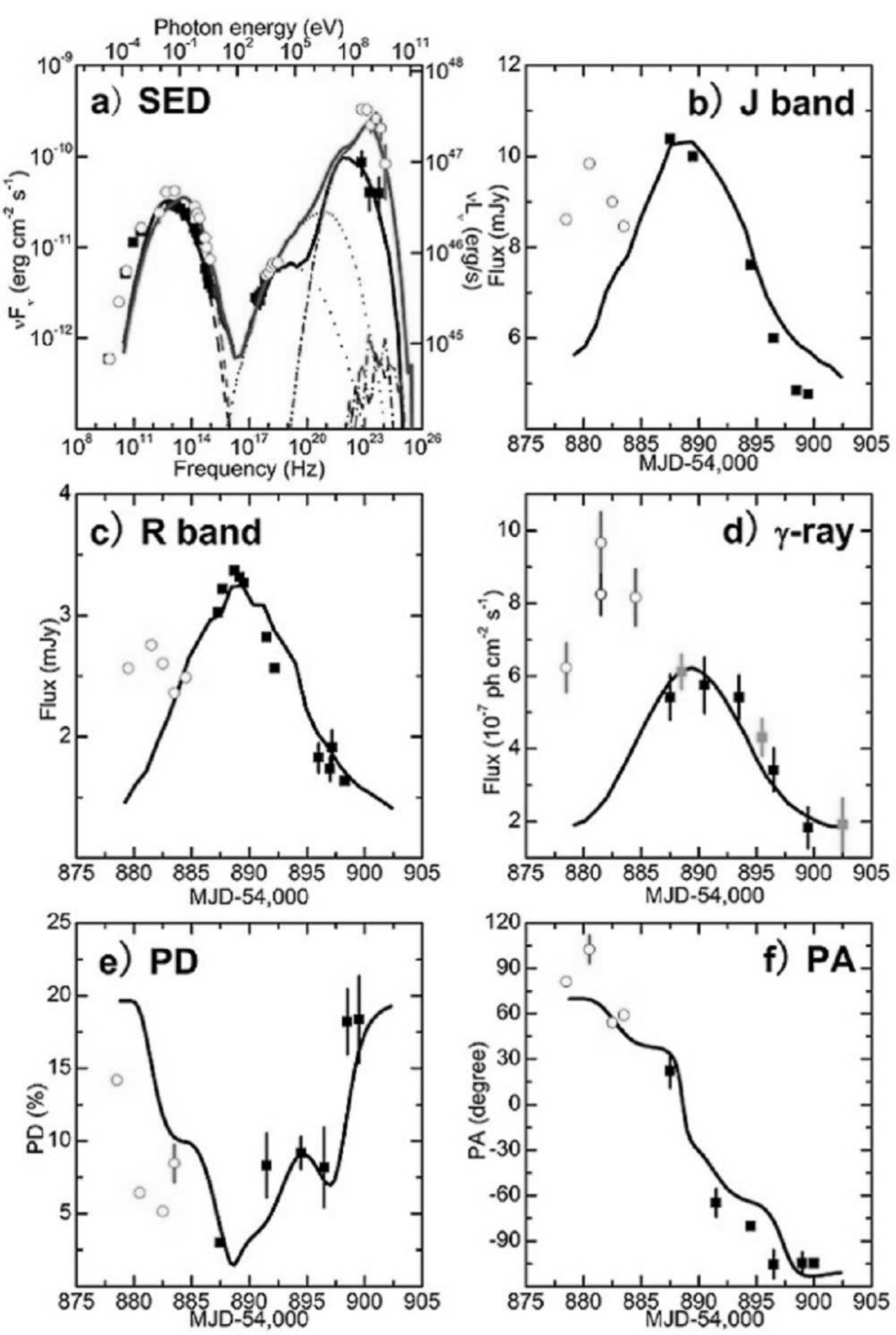}
\end{center}
\caption{Simultaneous fits to flux and polarization signatures of the FSRQ 3C279 during
the correlated PA swing + multiwavelength flare of 2009 \citep{Abdoetal2010}, using the
polarization-dependent shock-in-jet radiation transfer model of \cite{Zhangetal2014}.
(a) Snap-shot SEDs in quiescence and at the peak of the flare; (b -- d) flux light
curves in the IR (J-band), optical (R-band) and $\gamma$-rays (Fermi-LAT); (e) time-dependent
optical polarization degree; (f) time-dependent polarization angle, reproducing the observed
PA swing. From \cite{Zhangetal2015}.}
\label{3C279fits}
\end{figure}

In the framework of the Turbulent Extreme Multi-Zone (TEMZ) model \citep{Marscher2014}, 
the turbulence also reproduces variations of the linear polarization angle (PA), possibly 
including large-angle PA swings occasionally observed during FSRQ monitoring 
\citep[e.g.,][]{Marscheretal2010}. An alternative explanation for such PA swings,
correlated with multi-wavelength flares, was suggested by \cite{Zhangetal2014,Zhangetal2015},
in the context of an inhomogeneous shock-in-jet model. They have shown that such a model,
in which the emission region is pervaded by a helical magnetic field, naturally produces
PA swings of $180^o$ (or multiples thereof in the case of multiple disturbances propagating
through the jet) associated with multiwavelength flares, when carefully accounting for 
all light travel time effects, without the need for any asymmetric jet structures
(such as bends or helical motions of plasmoids along the jet). In \cite{Zhangetal2015},
this model was successfully applied to model, for the first time, snap-shot SEDs,
multiwavelength light curve {\it and} time-dependent optical polarization signatures
($\Pi$ and PA) of 3C279 during the correlated PA swing + $\gamma$-ray flare \citep{Abdoetal2010}
of 2009, in one coherent model (see Figure \ref{3C279fits}). In a follow-up study, 
this model was more self-consistently coupled to the results of MHD simulations of 
relativistic shocks in blazar jets in order to constrain, from first principles, the
particle energetics and magnetic-field topology \citep{Zhangetal2016}. 

The magnetic reconnection scenario also appears rather promising. Dedicated 
Particle in Cell plasma simulations \citep{SironiSpitkovsky2014} indicate 
that the combined action of acceleration at X-points and the subsequent Fermi-II 
like acceleration from the magnetic islands could provide an efficient way to 
rapidly accelerate electrons (and ions). A possible problem for this scenario 
is that BL Lac jets (in which most of the ultra-fast events have been recorded) 
appear to be weakly magnetized when the SED is reproduced by standard emission 
models \citep{TavecchioGhisellini2016}. 
Even if in the process of magnetic reconnection, the original magnetic 
energy density is reduced in the reconnection zones, much of the radiative energy
dissipation of accelerated particles is expected to occur in the remaining magnetic 
islands, in which the magnetization still has to be high for magnetic reconnection
to occur.

Another possible multi-zone model especially relevant for BL Lac jets is the 
so-called {\it spine-layer} model \citep{Ghisellinietal2005,TavecchioGhisellini2008}. 
In this picture, the jet is thought to be composed by two regions, an inner spine, 
moving with bulk Lorentz factor $\Gamma_s = 10$ -- $20$, surrounded by a slower 
($\Gamma _l \simeq 5$) layer. This kind of structure could result naturally 
from the jet acceleration mechanism itself \citep{McKinney2006} or due to
the interaction with the external environment \citep{Rossietal2008}. If both 
structures produce radiation, the energy density of the radiation produced by 
one component is relativistically amplified in the rest frame of the other 
component by the square of the relative Lorentz factor, 
$\Gamma_{\rm rel} = \Gamma_s \Gamma_l (1-\beta_s\beta_l)$. 
In this way the IC emission of the spine (which, having a much larger Lorentz factor, 
will preferentially dominate the emission at small viewing angles) receives a contribution 
from this kind of EC process. This structure could also be important in the possible 
production of high-energy neutrinos from BL Lac jets \citep{Tavecchioetal2014} and 
to unify BL Lacs with the misaligned parent population of FRI radio galaxies 
\citep{Chiabergeetal2000}.

\subsection{Location of the emitting region}
\label{location}

Therefore the location of the emitting region in the jet is still
quite hotly debated, especially for the case of $\gamma$ rays. The
problem is more acute for FSRQs for which the radiative environment of
the jet changes significantly along the jet. As we have pointed out
above, most of the models in the past assumed that the emission occurs
within the BLR radius. Indeed, this is consistent with the variability
timescale \citep[down to a few hours,][]{Tavecchioetal2010} and,
importantly, the large radiative energy density ensures a large
efficiency of the EC power, allowing one to minimize the total jet
power \citep{Ghisellinietal2014}. However there is independent
evidence that, at least occasionally, the emission occurs beyond the
BLR radius, possibly even at several parsecs from the center
\citep[e.g.,][]{Agudoetal2011}.

Important (almost model independent) evidence comes from the detection of VHE $\gamma$-ray 
photons from a handful of FSRQs (3C 279, PKS 1510-089, PKS 1222+216, PKS 1441+25, B0218+357) 
during active flaring states.  The discovery of FSRQ as VHE sources was quite surprising since, 
in the standard scenario, a huge opacity for gamma rays with energies above few tens of GeV is 
expected \citep[e.g.,][]{DoneaProtheroe2003,LiuBai2006}.  The UV radiation 
from the BLR (lines and continuum) is a target for $\gamma\gamma$ pair production, 
whose energy threshold is $E_{\gamma, \rm th} = 25 /(E_{\rm le}/10 \,{\rm eV})$ GeV. 
The optical depth starts to be important at an energy corresponding to the intense 
H Ly$\alpha$ line, $E_{\rm le} \sim 10$~eV, that is for $E_{\gamma} \sim 25$~GeV.  The detection 
of photons with $E > 30$~GeV therefore, especially in powerful sources -- for which the optical 
depth should be particularly large, naively increasing as $\tau\approx L_{\rm disc}^{0.5}$ -- 
immediately leads to the inference that the emission should occur beyond (or very close to) $R_{BLR}$ 
\citep{Paccianietal2014,BoettcherEls2016,Cooganetal2016}. Models for the VHE emission from 
FSRQs adopt this view, and generally assume that the observed high-energy radiation is 
produced through the IC scattering of the IR radiation field of the molecular torus 
\citep{Tavecchioetal2011,Aleksicetal2011bis,Paccianietal2014,Aleksicetal2014,Ackermannetal2014,Ahnenetal2015}. 

For quiescent states there is no evidence of VHE emission from FSRQs (although this conclusion is 
clearly influenced by sensitivity limits). Therefore, it is not clear whether the quiescent emission 
should also originate outside the BLR. 
Through the analysis of long-term {\it Fermi}/LAT data of bright FSRQs, \cite{PoutanenStern2010} 
claimed that the spectra show a well-defined and sharp break close to 6 GeV, which can be readily 
interpreted as induced by absorption through the interaction with the He II Ly$\alpha$ line/recombination 
continuum, expected from highly ionized clouds.  On the other hand, such a break could also be intrinsic to the emitted 
spectrum, arising from the combination of two components, namely, the Compton-upscattered disk and BLR radiation 
fields \citep{FinkeDermer2010}, or from Klein-Nishina effects and a curving electron distribution 
\citep{Cerrutietal2013}. Further investigations showed that the amount $\Delta\alpha_{\gamma}$ of the  
break is variable, with also the indication of a possible anti-correlation between the column density 
of the He II Ly$\alpha$ continuum and the gamma-ray flux \citep{SternPoutanen2011}. A possible way to 
interpret this complex phenomenology is to assume a stratified BLR, with lines of decreasing ionization 
produced at increasing distances from the central engine and with the emission region changing position 
with time \citep{Sternpoutanen2014}.


\subsection{A blazar sequence?}
\label{blazarsequence}

\begin{figure}[!htp]
\hspace{1.5truecm}
\includegraphics[trim = 0mm 5mm 0mm 0mm, clip,width=9cm,angle=0]{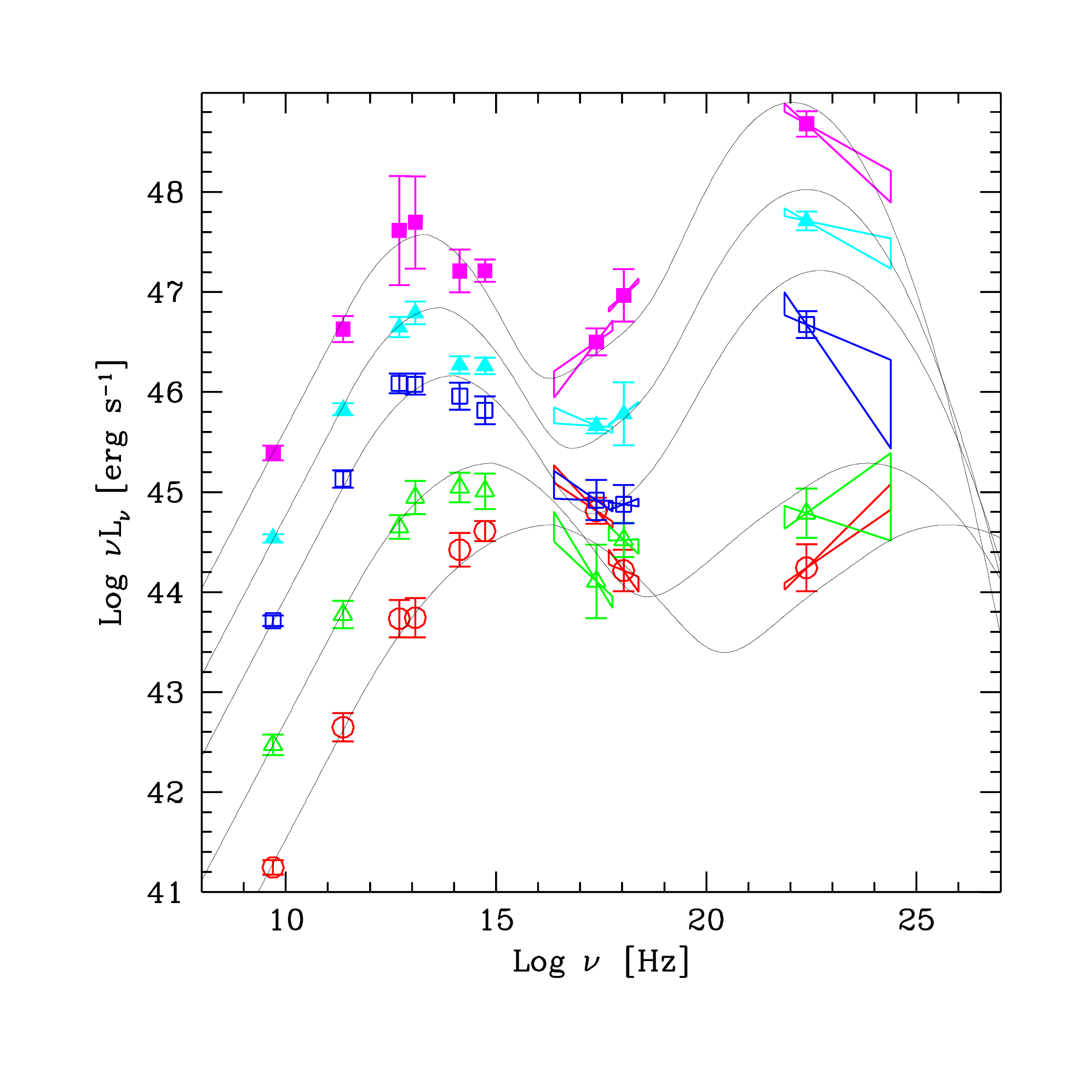}
\caption{The data-points show the average SED of blazars in bins of different radio (5 GHz) luminosity.  
The polynomial fitting curves (black lines) mark an inverse trend connecting the radio luminosity and 
the peak frequency of the two spectral component  (the {\it blazar sequence}). From \citealt{Donato2001}.}
\label{Figsequence}
\end{figure}

Soon after the discovery of the intense gamma ray emission from blazars, \cite{Fossatietal1998} reported 
the existence of a trend between the radio luminosity (at 5 GHz) of a blazar and the position of the two 
SED peaks (see Fig. \ref{Figsequence}). This trend was dubbed the {\it blazar sequence}.
According to the basic scheme, the SEDs of powerful sources (FSRQs) peak in the mm and MeV bands. Considering 
sources of lower luminosity, the two peaks progressively shift to higher frequency, reaching the X-ray band 
and the TeV band, respectively, in the case of the low-luminosity BL Lac objects. Together with this trend, one can also 
observe a decreasing {\it Compton dominance}, defined as the ratio of the luminosity of the high-energy bump 
with respect to the low energy one (see also \citealt{Finke2013}).

\begin{figure}[!htp]
\hspace{1.5truecm}
\includegraphics[trim = 0mm 5mm 0mm 0mm, clip,width=9cm,angle=0]{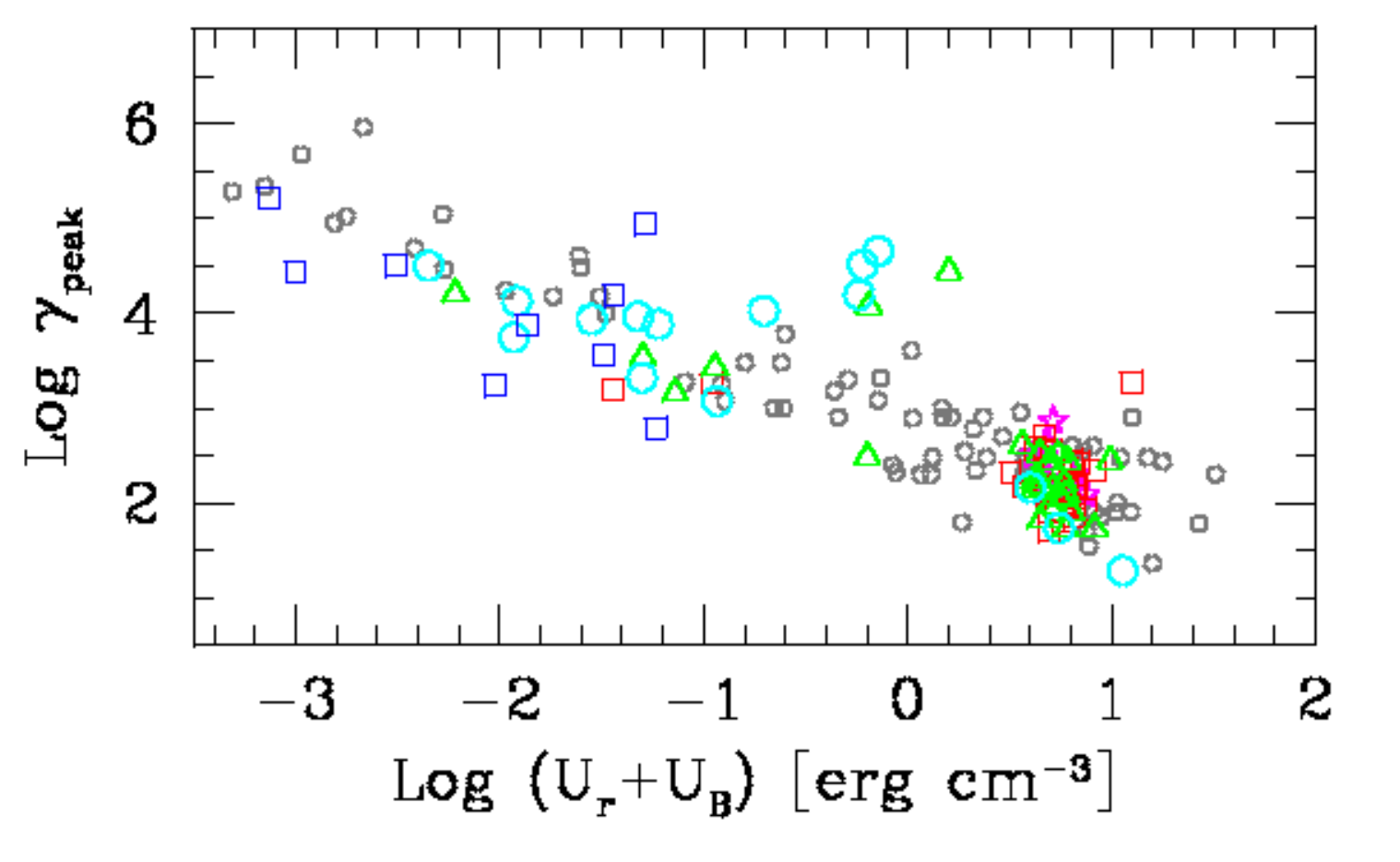}
\vspace*{0.5truecm}
\caption{Lorentz factor of relativistic electrons emitting at the SED peak versus the total energy density 
(magnetic plus radiative) for a sample of blazars. From \cite{Ghisellinietal2010}.}
\label{Figphyssequence}
\end{figure}

The SED modeling \citep{Ghisellinietal1998} shows that the {\it spectral} blazar sequence is accompanied by 
a correlation between two physical quantities, namely the energy (or the Lorentz factor) of the electrons 
emitting at the SED peaks $\gamma_{\rm p}$ and the sum of the energy density in magnetic field and radiation, 
$U = U_{\rm mag} + U_{\rm rad}$ of the form $\gamma_{\rm p} \propto U^{-1/2}$ (Fig.\ref{Figphyssequence}). 
This led \cite{Ghisellinietal1998} to heuristically propose that the main agent ruling the blazar sequence 
is the radiative cooling of the relativistic electrons. In fact, assuming a universal acceleration mechanism 
acting in all jets ensuring a characteristic acceleration rate per particle $\dot{\gamma}_{\rm acc}$, the 
equilibrium between acceleration gain and radiative losses $\dot{\gamma}_{\rm acc} = 
\vert \dot{\gamma}_{\rm cool} \vert \propto \gamma^2 U$ would naturally lead to the derived correlation.  
While the first work included only a few objects, with quite sparse data, subsequent studies using larger 
samples, more complete SED and better $\gamma$-ray data, confirmed the existence of the trend 
\citep{Ghisellinietal2010,GhiselliniTavecchio2015}. The above {\it cooling scenario} has been further 
extended in \cite{GhiselliniTavecchio2008}, to include scaling laws between black hole mass, AGN luminosity, 
jet power and BLR radius. In this scheme the ultimate parameter determining the blazar appearance could be 
the accretion rate onto the central supermassive black hole in Eddington units 
\citep{GhiselliniMaraschiTavecchio2009}.

Still not clear is the connection between the general trend shown by the sequence (peaks shifting towards lower 
frequency for increasing emitted power) and the behavior of single sources which, instead, generally follow the 
opposite pattern between quiescent and flaring states. A possibility is that the evolution is still dictated by 
the equilibrium between cooling and acceleration rate, with the latter increasing during active states. This 
seems to be the case in the well studied BL Lac Mkn 421 \citep{Mankuzhiyiletal2011}, for which the SED fits of 
different states reveal that the magnetic field in the emitting region and the energy of the electrons emitting 
at the SED peaks are related by the expected relation $B\propto \gamma^{-2}$. On the other hand, for the other 
prototypical VHE BL Lac, Mkn 501, the evidence suggests that only the electron energy is changing 
\citep{Tavecchioetal2001,Mankuzhiyiletal2012}. In this case a possibility is that the parameter regulating the 
SED is the acceleration efficiency. 

The actual existence of the blazar sequence is the subject of a lively debate. Most of the criticisms focus 
on the fact that observational biases could severely affect the selection of these sources, possibly introducing 
spurious trends. For instance, about $40 -- 50$~\% of the known BL Lac objects lack a reliable redshift 
measurement (made difficult by the weakness of their emission lines) and this could hamper the identification 
of far, powerful BL Lacs, which would patently violate the SED sequence. As a way to circumvent this limitation,
\cite{Finke2013} suggested to use the Compton dominance instead of luminosity as the quantity to correlate 
with the SED peak frequencies, which does not require the knowledge of redshifts (with only a negligibly small
error introduced by a lack of knowledge in the redshift of the peak frequencies). Using this new correlation
discriminant, he found renewed evidence for the existence of a sequence, in agreement with the original
\cite{Fossatietal1998} sequence for BL Lac objects, especially at high peak frequencies, but essentially
no correlation for blazars with synchrotron peak frequencies below $\sim 10^{14}$~Hz, with Compton dominance
parameters spreading over $\sim 3$ orders of magnitude without any correlation with the synchrotron peak frequency. 
The efforts devoted to investigate the sequence at low luminosities also start to show possible deviations from 
the expectations \citep[e.g.,][]{CaccianigaMarcha2004,Padovanietal2007,RaiteriCapetti2016}, although the situation 
is particularly delicate since low luminosity non-blazar AGNs, misaligned (hence less beamed) blazars or 
blazars with (relatively) low mass black holes could ``pollute" the samples. These criticisms have been 
condensed by \cite{Giommietal2012} \citep[see also][]{Giommietal2013,GiommiPadovani2015} into a simplified 
view aiming at demonstrating that, even without any real trend in the intrinsic SED, selection effects alone 
can be responsible for the observed spectral sequence. 

It is important to remark that, in discussing these issues, one should always clearly distinguish between 
the (observational) blazar sequence, in the sense of \cite{Fossatietal1998}, and the (theoretical) cooling 
scenario (and its extensions). In fact, even if the spectral sequence could be affected by selection effects, 
the latter is a more general physical scheme which, in fact, can accomodate or even envisages the existence 
of possible outliers from the canonical blazar sequence 
\citep[e.g.,][]{GhiselliniTavecchio2008,Ghisellinietal2012,Ghisellinietal2013}.  Note also that, according 
to the general scenario developed in \cite{GhiselliniTavecchio2008}, the observed sequence could just be the 
``tip of the icerberg" of the blazar population, representative only of the most massive and powerful blazars.  
Sources with smaller black hole masses or slightly misaligned jets are expected to produce low-luminosity jets 
with, however, a SED shape similar to that of the  powerful FSRQ, thus appearing as outliers from the blazar 
sequence. In this sense the observed sequence could just be the upper edge of an envelope of progressively 
misaligned blazars \citep{Meyeretal2011}.

\section{X-Ray Binaries --- Microquasars}
\label{microquasars}  

Over the last decades, it has become clear that many aspects of
accretion physics are mirrored across the black hole mass scale,
irregardless of the size difference and accretion environment.  The
extent of this 'mass-scaling' suggests the relative importance of
mechanisms within the inner accretion flow rather than those dictated
by fueling and/or outer disk geometry, and gives clues about the
nature of jet launching.  Although our theoretical understanding still
 lags the quality of the precision, broadband data in many cases, we
 are starting to find empirical trends that allow an urgently needed
 reduction in the number of free parameters.    

 In isolation, the modeling of jets and particle acceleration in XRBs
 is subject to the same uncertainties and avalanch of free parameters
 as discussed in the above Sections for AGN.  However because XRBs
 evolve on much faster timescales, they offer a unique perspective,
 where an individual can traverse a wide variety of behaviors that
 only large samples of AGN can equally populate.  This has opened a
 new window on one of the more persistent problems to solve in jet
 physics: the coupling between bulk dynamical properties (or
 macrophysics), and the microphysics of processes like particle
 acceleration and radiation.  In this Section we will summarize some
 of the recent progress in this area and its extension to AGN.

 XRBs come in several flavors, high and low mass (with respect to the
 companion star), and with black holes or neutron stars as the compact
 object.  In order to compare with AGN we will be limiting ourselves
 to discussing black hole XRBs.  The complicated phenomenology of XRBs
 has been covered in several papers and reviews \citep[see
 e.g.,][]{McClintockRemillard2006,Fender2006, Belloni2010} to which we
 refer the interested reader.  Here we will only summarize the most
 relevant properties, in order to focus the discussion on comparisons
 with jetted AGN with a focus on breaking model degeneracy and better
 understanding particle acceleration.  

\begin{figure}
\centering
\includegraphics[width=\columnwidth]{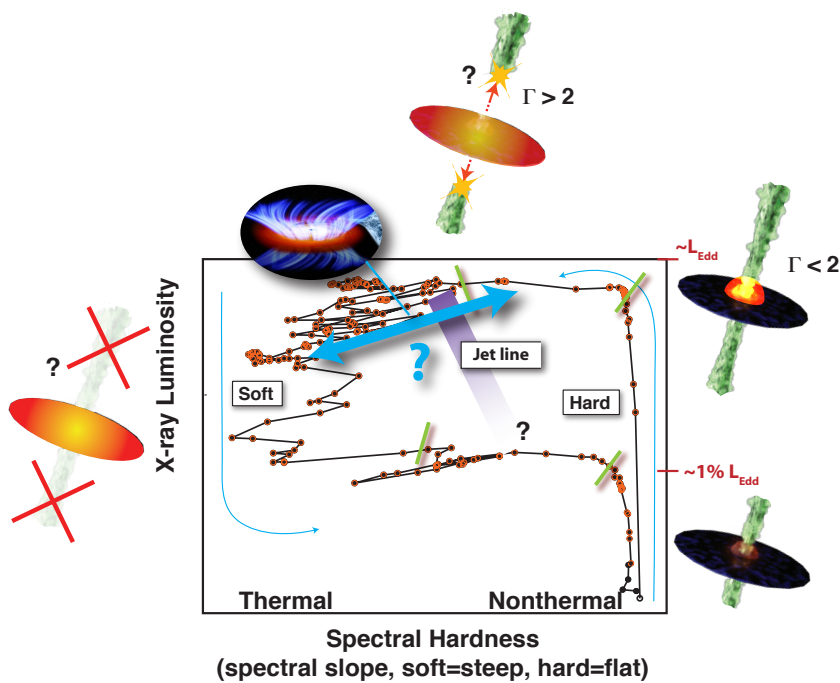}
\caption{The hardness-intensity diagram (HID) for an outburst of XRB
  GX339$-$4, with schematics of the corresponding inflow/outflow
  configuration per state.   Only the two canonical states (hard and soft)
  are indicated.}
\label{fig:hid}
\end{figure}

Low mass XRBs (LMXBs) are black holes with a main-sequence companion
star whose mass is smaller than the compact object, typically on the
order of a solar mass.   The star fills its Roche lobe and thus feeds
material to the black hole via its L1 Lagrangian point, which leads to
a well defined entry point for material to populate an accretion disk
in the orbital plane.   LMXBs go through transient outbursts, likely
driven by thermal instabilities in the accretion disk
\citep[e.g.][]{Lasota2001}, which initiate a sequence of distinct
states, originally classified in terms of X-ray and spectral timing
properties.   Fig.~\ref{fig:hid} shows a schematic view of this
pattern of cyclic behavior,  using actual data from the Galactic LMXB 
GX339$-$4, which has been a great boon for our
understanding because it experiences a full outburst cycle
approximately every 1-2 years.  The figure shows the outburst
pattern in a commonly used phase space of X-ray
luminosity (related to, not necessarily linearly) the accretion rate
$\dot{M}$ onto the black hole, and the X-ray spectral hardness.   The
latter indicates the relative dominance of thermal (soft spectrum)
versus nonthermal (hard power law) processes.   This type of diagram
is referred to a hardness-intensity (HID) diagram.

During a typical outburst, a LMXB rises out of quiescence (off the
plot in the lower right corner in Fig.~\ref{fig:hid}), a state which
is quite difficult to study by definition because of the very low
luminosity.  The weakest sources seen in quiescence so far are
XTE~J1118$+$480, A0620$-$00, and Swift J1357.2$-$0933, all with
$L_{\rm X}\leq 10^{-8} L_{\rm Edd}$
\citep[e.g.][]{Galloetal2007,Galloetal2014,Plotkinetal2016}.  At some
point the rising X-rays trigger a monitoring instrument such as
\textit{SWIFT} or \textit{MAXI} (in the past it was typically
\textit{RXTE}) and these facilities then send out alerts, allowing
further multiwavelength coverage.  At the point that a source is
bright enough to trigger facilities, it is already in the so-called hard state.
This state is associated with compact, steady jets which are
synchrotron self-absorbed, resulting in the classical flat/inverted
spectrum associated also with the compact cores of AGN
\citep{BlandfordKoenigl1979}.  The hard state is the longest lasting
of the LMXB states and, while quiescence has been argued to be a
separate state, the latest observational evidence supports the
persistence of jets smoothly down into quiescence, as well as other
properties \citep[see, e.g.,][]{Plotkinetal2015}.

As the source luminosity increases (assumedly driven by a rising
accretion rate), somewhere around $\sim0.5L_{\rm Edd}$ the spectrum
begins to soften and the source moves counter-clockwise around the
upper right corner in the diagram, beginning a transition into a state
often referred to as the Hard Intermediate State (HIMS).  These
transitions are harder to categorize based on radio or X-ray spectra,
but are clearly demarcated via changes/features in the power density
spectra \citep[e.g.,][]{Belloni2010}, that we will not discuss in
detail here, but that include the presence of quasi-periodic
oscillations (QPOs) and various types of broadband variability.  The
hard-to-HIMS transition occurs very quickly and within a day or days,
the jets are seen to radically transform from steady flow to a
transient state with brighter, discrete ejecta at a distinct point
often referred to as the ``jet line'' in the HID.  The jets appear now
as discrete optically thin blobs moving with superluminal motion; it
was on this state that the original argument for the nickname
``microquasar'' was based \citep{Mirabel1994}.  After the flaring
ends, the source is in the Soft Intermediate State (SIMS) and the jets
essentially deconstruct: the radio and IR synchrotron radiation
vanishes and the source proceeds into the left side of the diagram,
into a soft thermal state that is almost entirely disk dominated, with
a multi-temperature black body.  As the luminosity decreases
(assumedly along with $\dot{M}$), the spectrum begins to harden again
and the jets reform, at a luminosity around an order of magnitude or
more below the hard-to-soft transition.  This hysteresis effect is not
well understood, but it seems clear that something besides the
accretion rate, likely magnetic field topology, is an important
driver.  With renewed steady jets, the hard state recedes down into
quiescence and the outburst is over.  A typical outburst can last from
months to $\sim$ a year.

In contrast to LMXBs, HMXBs tend to be persistent, with Cyg X-1 seen
as the ``prototypical'' source of its class.  Cyg X-1 does show state changes, but
with a much lower range in luminosity, and because it does not go into
quiescence it never undergoes the full hysteresis loop
\citep[e.g.][]{Gleissneretal2004,Wilmsetal2006}.   Mostly it varies
between soft and hard states.   In contrast to LMXBs, only a few HMXBs
are known in our Galaxy \citep{Tetarenkoetal2016}.

The two states that can be associated with particle acceleration via
the presence of jets are thus the hard and HIMS.  Interestingly, the
discovery of jets in XRBs came over 30 years after their discovery as
a class via X-ray observations, at which point a paradigm based
entirely on accretion disk phenomena was already well established.
The X-ray hard state SED is typically comprised of a nonthermal power law with
$\Gamma \sim 1.5-1.8$, and an exponential cutoff around 100 keV.  Sometimes
there are also weak reflection features such as an iron fluorescent
line and a ``Compton bump'' leading to hardening above $\sim 15$~keV
\citep{LightmanWhite1988,Fabianetal1989}.  The hard state SED can be
explained in terms of a moderately optically thick ($\tau\lesssim1$),
static region of hot electrons existing near the black hole and disk.
The existence of this region, named the corona, was inferred in order to explain the hard
X-ray power-law via unsaturated inverse Compton scattering.
In this scenario, the corona upscatters the weaker disk thermal
photons (not usually visible in the spectrum, but assumed present
because of the clear thermal disk signature in the soft state).  This
type of scenario seems natural, because the temperature that gives
both the spectral index (for some optical depth) and cutoff is
consistent with the Virial temperature for a compact region within 10s
of $r_g$ from the black hole \citep[see,
e.g.,][]{Titarchuk1994,MagdziarzZdziarski1995}.

As is common in astrophysics, however, increasing precision in the
observations tends to provide challenges.  More extensive monitoring
after the launch of \textit{RXTE} revealed significant variability in
the cutoff over relatively short timescales
\citep[e.g.,][]{Rodriguezetal2003,Wilmsetal2006}, suggesting a more
dynamic region than originally envisioned.  Another complication was
the relatively weak reflection signatures in many observations.  In
order to reduce the amount of coronal emission intercepted by the
cooler disk, many variations of geometry were developed, including a
recessed disk, a patchy corona, magnetic flares, etc. \citep[see,
e.g.,][]{Haardt1997,MerloniFabian2002}.  One paper in particular
suggested that a magnetic corona could be accelerated to moderately
relativistic speeds away from the disk, thus providing an explanation
for the relatively weak reflection signatures in Cyg X-1
\citep{Beloborodov1999}.  This latter scenario is especially relevant
in the context of integrating jets into the existing coronal picture.

The problem is that relativistic jets need dynamically important,
ordered magnetic fields theading the disk near the black hole, in
order to launch \citep[e.g.][]{McKinney2006,BeckwithHawleyKrolik2008}.
The need for these fields creates two main complications to the
original picture: first, the coronal electrons will want to follow the
magnetic fields lines and the static picture breaks down.  Secondly,
the introduction of a strong magnetic field means that the electrons
will also radiate (and thus cool) via synchrotron emission, and those
photons can also be inverse Compton upscattered (SSC) along with those
from the disk.  

The presence of jets thus requires some rethinking of the original
picture, and highlights the need to find a self-consistent link
between the three types of flow: inflowing disk, corona, and
outflowing jets.  The problem is that multiple regions introduces
degeneracy for the interpretation of the hard X-ray spectrum, which
can now be from a thermal corona, a hybrid thermal/nonthermal corona,
an isolated nonthermal corona or a magnetized corona comprising the
base of the jets, as well as direct synchrotron emission
\citep{MarkoffNowakWilms2005}.  In the meantime, more recent studies
of reflection suggest that the data are more consistent with a source
of disk illumination that lies on the axis above the disk, and changes
its height \citep[``lamppost'' reflection, see,
e.g.,][]{MiniuttiFabian2004,Dauseretal2013}.  This idea further
supports a hybrid corona/jet base, where the effect of even mildly
relativistic acceleration of the jets away from the disk means that
IC/SSC from the jet base will dominate over synchrotron emission in the
reflected component, unless particle acceleration starts within 10s of
$r_g$ of the black hole \citep[see, e.g.,][]{MarkoffNowak2004}.

In an attempt to rule out one or more of the various hard state
emission scenarios, \cite{Nowaketal2011} carried out an extensive,
simultaneous campaign on Cyg X-1 using \textit{Suzaku}, \textit{RXTE}
and \textit{Chandra-HETG}.  Instead, they found that all main
(Compton-dominated) scenarios provided equally good statistical
descriptions of the new data sets.  On the other hand, the fact that
XRB jets clearly emit synchrotron radiation up to the mm/IR bands,
depending on the observation \citep[e.g.][]{Fender2006}, presents a
way to probe the connection between the components.  For instance,
coordinated multiwavelength observations can help to isolate the
jet-related components, as well as build a picture of the jet
properties in XRBs compared to AGN (content, particle acceleration,
etc.).

\begin{figure}
\centering
\includegraphics[width=\columnwidth]{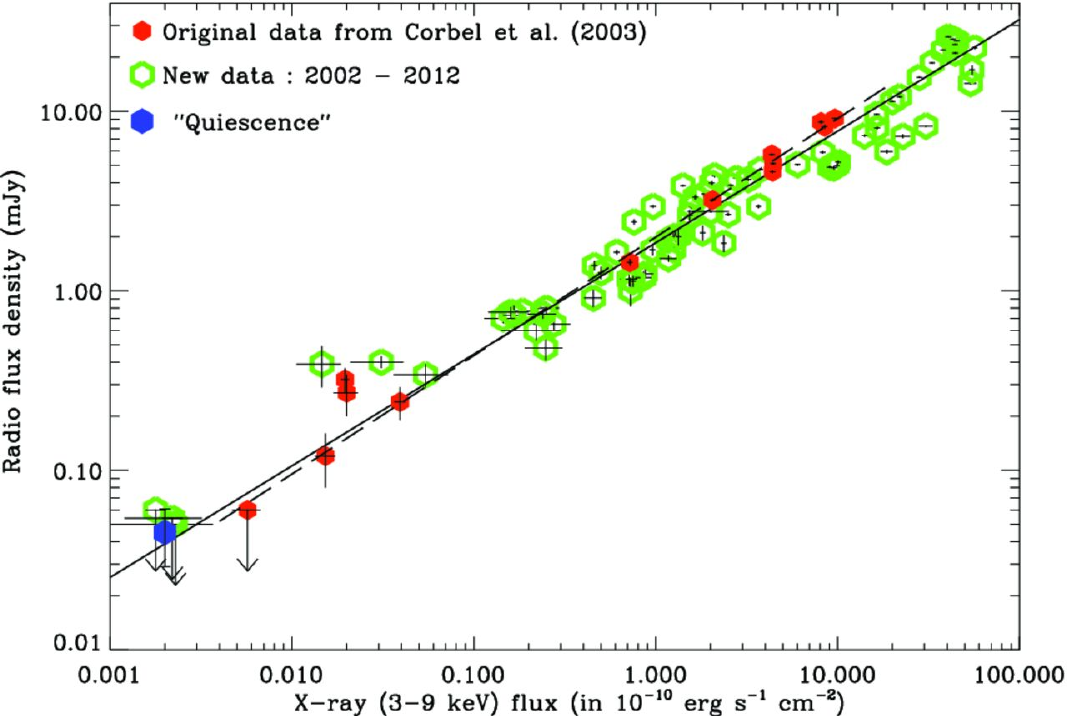}
\caption{Radio vs. X-ray correlation in the hard state of GX~339$-$4,
  with the original correlation shown in red from
  \cite{Corbeletal2003}, and the newer outbursts.  The dashed line
  illustrates the original radio/X-ray correlation from the 1997–1999
  period, while the solid line corresponds to a fit to the new whole
  sample. The blue point shows the source close to
  quiescence.  From \cite{Corbeletal2013}.}
\label{fig:corbel13a}
\end{figure}

Thus in parallel with some of the developments in ideas about the
corona, the era of coordinated, quasi-simultaneous observations of
XRBs really began to take off at the start of this century,
particularly between the radio and X-ray bands \citep[but see
also][]{Motchetal1982}.  Soon after, a tight, nonlinear correlation
was discovered between the radio and X-ray luminosities of GX339$-$4,
holding over many orders of magnitude, and repeated during different
outbursts \citep[Fig.~\ref{fig:corbel13a} and, 
e.g.,][]{Corbeletal2000,Corbeletal2003,Corbeletal2013}.  This
correlation was soon found to extend to other sources
\citep[e.g.][]{GalloFenderPooley2003}, as well as between the IR and
X-ray bands \citep{Russelletal2006}.

Because the different configurations of disk vs jet dominance in the
HID are also seen among the various AGN classifications, the idea soon
arose that AGN may be undergoing XRB-like state changes, just over
much longer timescales.  This idea of trying to ``pair''  XRB states
to AGN classes, often referred to as 'mass scaling' in accretion, has
been the topic of much exploration in the last $\sim13$ years.  The
most successful pairings to date involve the longer-lasting hard and
soft states.  For instance, there is a compelling mass-dependent
scaling between the power spectra of the soft/intermediate state and
that of Seyferts \citep[e.g.,][]{McHardyetal2006}.  However, in the
context of jets and particle acceleration, the most interesting
scaling is the so-called Fundamental Plane (FP) of black hole
accretion.

The FP is an empirical relation, essentially a 2D plane linking all
sub-Eddington, jetted black hole types in the 3D space of X-ray
luminosity, radio luminosity and mass.  After the discovery of the
radio/X-ray correlation in individual XRBs, it was quickly realized
that this correlation represents the ratio of the efficiencies for the
two radiative processes.  For instance \cite{Markoffetal2003}
considered the scaling in the context of synchrotron emission from the
jets.  Using a Blandford-K\"onigl-like jet model \citep{BlandfordKoenigl1979}, 
one can show that the self-absorbed, flat/inverted radio-through-IR part of 
the spectrum will scale as $\sim \dot M^{17/12-2/3\alpha_{\rm RIR}}$, where
$\alpha_{\rm RIR}$ is the spectral index of the self-absorbed part of
the jet and is close to 0 \citep[see also][]{FalckeBiermann1995}.  The
measured slope of the correlation is
$L_{\rm X} \propto L_{\rm R}^{0.6-0.7}$ \citep{Corbeletal2013}, which
constrains the dependence of $L_{\rm X} \propto \dot M^{2-2.3}$,
ruling out radiatively efficient processes for the hard X-ray
spectrum.  This result was generalized by \cite{HeinzSunyaev2003},
expressing all physical scales in terms of $M_{\rm BH}$ (via $r_g$)
and all powers in terms of
$\dot m \sim L_{\rm Edd}/(c^2) \propto M_{\rm BH}$.  This mass scaling
effectively introduces a normalization term into the radio/IR/X-ray
correlations.  In other words, if compact jets and accretion disks are
self-similar structures across the immense mass and power scales seen
by ``hard state-like'' black holes, then all black holes should show a
similar coupling of X-ray and radio luminosity as a function of
relative accretion rate, once ``corrected'' by the
mass.

\begin{figure}
\centering
\includegraphics[width=\columnwidth]{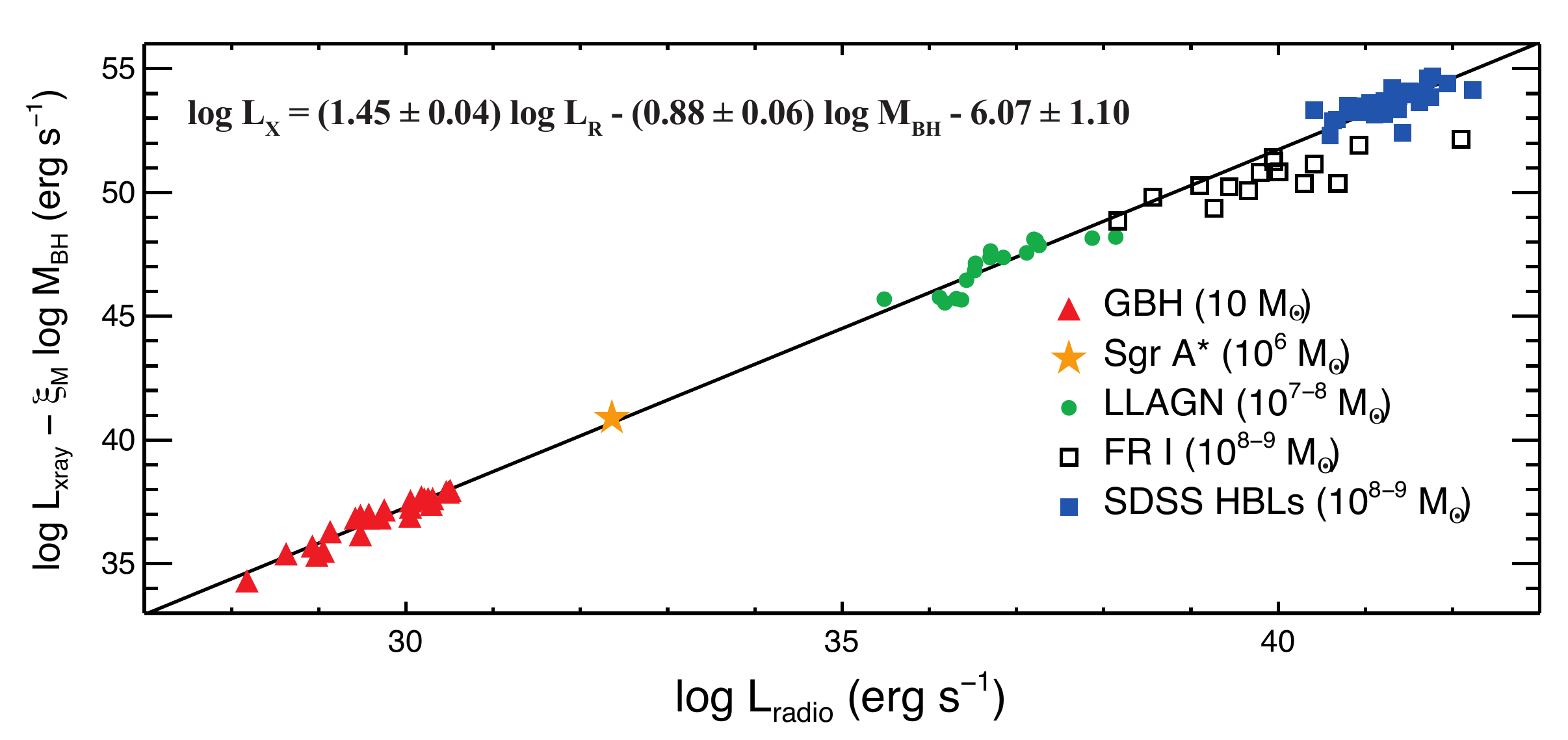}
\caption{A projection in mass of the Fundamental Plane for black hole
  accretion onto the radio and X-ray luminosity plane, using a
  Bayesian regression technique on data from hard state XRBs and a
  sample of sub-Eddington accreting AGN \citep{Plotkinetal2012}.  }
\label{fig:fp}
\end{figure}

This idea was explored and tested empirically by two different groups,
using large samples of AGN
\cite{MerloniHeinzDiMatteo2003,FalckeKoerdingMarkoff2004}.  If the AGN
follow a similar correlation as hard state XRBs over much longer
timescales, this trend would be revealed statistically by looking at
snapshots of a large number of objects.  Both groups, using different
approaches, found that such a correlation indeed seems to extend to
AGN, and in the subsequent years more discussion and exploration of
the statistics has ensued.  A good summary can be found in
\citet[]{Plotkinetal2012}, as well as the most robust derivation of
the FP, for the first time also including BL Lacs (see
Fig.~\ref{fig:fp}).  While there is still some debate over exactly
which AGN classes fit on the FP, it is pretty clear that
sub-Eddington, steady jetted sources such as 
Low Louminosity AGN (LLAGN) or Low-Ionization Emission-Line Regions (LINERs),
FRIs and BL Lacs do, and perhaps also Narrow-Line Seyfert-1 (NL Sy1)
with radio detections \citep[e.g.][]{Gueltekinetal2009}.

The extension of the broadband correlations found in XRBs to AGN has
therefore opened up a new pathway to study the disk/jet coupling in
black holes.  The X-rays probe the conditions near the black hole, in
a corona or near the launchpoints of the jet,  and its intimate
connection to the transfer of energy into accelerated particles
further out in the jets, responsible for the
radio/IR parts of the FP correlation.  

Having established that mass-scaling seems to hold in the disk/jet
coupling, we can now consider how to use it in order to  better understand the power
and acceleration properties of the jets in the hard state (for all
black hole masses).

\subsection{Origin of the high-energy emission}
\label{heorigin}

Returning to the question of the origin of the hard state emission,
the degeneracy between synchrotron and inverse Compton processes (and
for the latter, also between different contributing photon fields)
exists at both ends of the mass range.  In XRBs, the lower Lorentz
factor of the jets limits EC to two contributions: the accretion disk
or the companion star.  In LMXBs the companion star is not bright
enough to compete with the disk, reducing the possible source of the
X-rays to direct synchrotron radiation, SSC emission, EC of disk
photons, or some combination.  In most scenarios all of these could
give the correct scalings for the FP, thus the next step is to move
beyond correlations and look at modeling the broadband SEDs.

\begin{figure}
\centering
\includegraphics[width=\columnwidth]{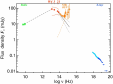}
\caption{Average dereddened SED of GX~339$-$4, including WISE data
  points for the first time (in red). Red curves represent the envelope 
  of extreme variations over 13 WISE epochs. The near-infrared and 
  optical/ultraviolet (OUV) points are plotted in orange and radio 
  in green, and the break from optically thick to optically thin is 
  apparent, see \citep{Gandhietal2011}. }
\label{fig:gandhi}
\end{figure}

The original suggestion that synchrotron could contribute
significantly to the hard X-ray spectrum in XRBs came from
\cite{Markoffetal2001}, for one of the first quasi-simultaneous
broadband (radio through X-ray) SEDs compiled \citep{Hynesetal2000}.
The hard X-ray spectrum shows little to no reflection, nor a high
energy cutoff, in contrast to what would be expected for a thermal
corona model.  Furthermore, jets can account for all the nonthermal components
in the spectrum, as long as the particle acceleration initiates
somewhat offset from the black hole, at around $50r_g$.  This
requirement of having a transition from primarily thermal plasma in
the jets \citep[similar to Sgr A*, see, e.g.,][]{Markoffetal2001b} to
a region where a powerlaw of particles is accelerated implies a break
in the spectrum where this enhanced emission transitions from
optically thick to thin.  This feature is not always possible to
detect because it often falls in the IR band where the companion star
and/or disk can contribute, but it can be seen explicitly in GX339$-$4
\citep[see Fig.~\ref{fig:gandhi}, and, e.g.][]{CorbelFender2002}, and
detected or constrained in many other observations
\citep{Russelletal2010, Russelletal2013, Koljonenetal2015}.

We will come back to the break in Section~\ref{breaks}, but its
location in the SED provides a hard limit on how much synchrotron
emission can contribute to the hard X-ray spectrum.  Unlike the SED
for XTE~J1118$+$480, most hard state XRBs, particularly at higher
luminosity, do show evidence for some reflection and thus a contribution in
SSC/EC from the corona, or jet base.  However synchrotron emission can
still contribute at the $\sim10\%$ level, which has implications for
particle acceleration in XRBs.  Even when the break is not visible,
there are independent methods to gauge the jet synchrotron
contribution.  For instance \cite{Russelletal2010} studies the
evolution of an outburst in the XRB XTE~J1550$-$564 in three IR bands.  During
the decline from outburst in the disk-dominated soft state, a clear
exponential decay is visible.  Upon transition to the hard state, an
excess emission above this monotonic decrease appears, corresponding
to the jets re-activating.  Extracting this excess allows an
independent determination of the jet synchrotron contribution to the
IR band, and its spectral index, which is found to match that of the
X-rays in both slope and normalization.

Another independent constraint on the contribution and geometry of the
jets comes from fast IR/optical and X-ray variability studies.  One of
the first examples of such a spectro-timing study is presented for
XTE~J1118$+$480 in \cite{Hynesetal2003}, where they extract the
broadband SED of the rms variability in IR, UV and X-ray, and find it
follows a $\nu^{-0.6}$ powerlaw, consistent with optically thin
  synchrotron emission.  More recently, \cite{Kalamkaretal2016} detect
  the first IR QPO during a hard state of GX339$-$4 (at half the
  frequency of the X-ray QPO, and the same frequency as that found in
  optical/UV/Xray from Hynes et al. 2003 for XTE~J1118), as well as
  strongly correlated sub-second variability, with the IR lagging the
  X-rays by $111$ms.  The direction of the lag, as well as brightness
  temperature and size arguments, effectively rules out reprocessing
  or hot-inflow models, similar to argumentation in Hynes et al. 2003.
  The timescale for the lag corresponds to a light-travel distance of
  $3700 r_g$ for $\sim6 M_\odot$, consistent with the distance of the
  IR-emitting region based on the location of the break in the
  spectrum of \cite{Gandhietal2011} (Fig~\ref{fig:gandhi}).  The first
  detection of a QPO in the IR further suggests a picture where
  oscillations rooted in the accretion flow are conveyed with the
  plasma into the jets, perhaps via ordered magnetic field lines.

Of all microquasars,  Cygnus X-1 is the most extensively observed.  
This source has been the target of  monitoring campaigns that allowed 
to estimate the parameters of the binary and to obtain detailed spectra 
at all wavelengths. Cygnus X-1 is located at $1.86~\rm{kpc}$ from Earth. 
A high-mass stellar companion of spectral type O9.7 Iab and mass \mbox{$\sim 20~M_{\odot}$} 
and a black hole of \mbox{$14.8~M_{\odot}$} \citep{Oroszetal2011} form the binary system. 
 
In the X-ray band Cygnus X-1 switches between the typical hard and soft 
states of X-ray binaries. The soft state is characterized by a blackbody 
component of \mbox{$kT \lesssim 0.5~\rm{keV}$} from an accretion disk, and a 
soft power-law with spectral photon index \mbox{$\Gamma \sim 2 -3$}. The source, 
however, spends most of the time in its hard state, in which the spectral 
energy distribution is well described by a power-law of spectral index \mbox{$\Gamma \sim 1.7$} 
that extends up to a high-energy cutoff at \mbox{$\sim 150$ keV} 
\citep[e.g.][]{Doveetal1997, Poutanenetal1997}. The origin of this power-law is the  
Comptonization of disk photons by thermal electrons in the hot corona that partially 
covers the inner region of the disk. The detection of a Compton reflection bump and 
the Fe K$\alpha$ line  at $\sim6.4$ keV support the presence of the corona during the 
low/hard state.   Additionally, intermediate spectral states have also been reported 
\citep{Bellonietal1996}. 

Persistent and transient jets have been resolved at radio wavelengths in Cygnus X-1 
during the hard state \citep[e.g.][]{Stirlingetal2001, Rushtonetal2011}. The outflow 
is extremely collimated \citep[aperture angle $\sim 2^\circ$,][]{Stirlingetal2001} and 
propagates at an angle of $\sim 29^\circ$ with the line of sight \citep{Oroszetal2011}. 
The radio emission is modulated by the orbital period of the binary because of absorption 
in the wind of the companion star \citep[e.g.][]{Brocksoppetal2002}.

Cygnus X-1 is one of the two confirmed MQs that is a gamma-ray source.\footnote{The other 
one is Cygnus X-3. }  The first detection of soft gamma rays up to a few MeV was achieved 
with the instrument \emph{COMPTEL} aboard the \emph{Compton Gamma-Ray Observatory} 
\citep{McConnelletal2002}. Emission in the same energy range was later observed with 
\emph{INTEGRAL} \citep[e.g.][]{Laurentetal2011}. The  \emph{INTEGRAL} detections 
represented a breakthrough since it was found that the \mbox{$\sim 400$ keV - 2 MeV} 
photons were highly polarized. 

At higher energies Cygnus X-1 is fundamentally a transient source on timescales of 1-2 d. 
Episodes of gamma-ray emission have been detected with the satellite \emph{AGILE} between 
100 MeV and a few GeV in the hard state (\citealt{Sabatinietal2010},  and marginally 
during the hard-to-soft transition \citep{Sabatinietal2013}. The analysis of 
\emph{Fermi}-Large Area Telescope (LAT) data at 0.1-10 GeV revealed weak flares (three 
of them quasi-simultaneous with \emph{AGILE} detections; \citealt{Bodagheeetal2013},  
and weak steady emission \citep{Malyshevetal2013}.

Finally, Cygnus X-1 has been observed in the very high energy band ($\geq 100$ GeV) 
with the Major Atmospheric Gamma Imaging Cherenkov (MAGIC) telescope  during the hard 
state \citep{Albertetal2007}. During inferior conjunction a flare (duration of less than a 
day, rising time $\sim 1$ h) was likely detected with a significance of  $4.0\sigma$  ($3.2\sigma$ 
after trial correction).  Only upper limits could be obtained for the steady emission. 

The broadband spectral energy distribution (SED) of Cygnus X-1 in the hard state displays 
several components. The radio emission is synchrotron radiation of relativistic electrons 
accelerated in the jets. This component is observed up to its turnover at infrared frequencies, 
where the stellar continuum takes over. The emission of the disk/corona dominates up to the 
hard X-rays, but the origin of the MeV tail observed with \emph{COMPTEL} and \emph{INTEGRAL} 
is still disputed. Its high degree of polarization (>65\%) suggests this component is emitted in 
an ordered magnetic field such as is expected to exist in the jets, a result supported by 
the fact that the polarized X-ray emission is only clearly detected during the hard 
state \citep{Rodriguezetal2015}. In this scenario the MeV tail would be the cutoff of the 
jet synchrotron spectrum, see for example the fits to the data obtained by  \cite{Malyshevetal2013},  
\cite{Zdziarskietal2012}, \cite{Zdziarskietal2014}, and \cite{Zhangetal2014}. An alternative 
model was introduced by \cite{Romeroetal2014}, where the MeV tail is synchrotron radiation of 
secondary non-thermal electrons in the corona. This model predicts significant polarization 
also during intermediate spectral states, something that cannot be presently ruled out from 
the data \citep{Rodriguezetal2015}.

All known gamma-ray binaries host a high-mass donor star, a fact that points to a fundamental 
role played by the stellar wind and/or  radiation field in the mechanisms that produce the 
high-energy photons. In leptonic models for MQs gamma rays are produced by inverse Compton (IC) 
scattering of the stellar radiation off relativistic electrons, whereas in hadronic models 
gamma-rays are generated by the decay of neutral pions injected in the interactions of non-thermal 
protons in the jets with cold protons of the stellar wind \citep{Romeroetal2003, Romeroetal2005}.  

The inclusion of relativistic protons in the jets brings about a feature absent in purely leptonic 
models, namely the production of secondary particles (neutrinos, electron-positron pairs, muons, pions) 
in high-energy hadronic interactions. The cooling of charged secondaries may contribute to the radiative 
spectrum of the jets. In Cygnus X-1 the effects of the impact of the jets in the interstellar 
medium suggest that they carry a significant amount of kinetic energy in cold protons 
\citep{Galloetal2005, Heinz2006}.

\begin{figure}%
\centering
\includegraphics[width=1.1\columnwidth]{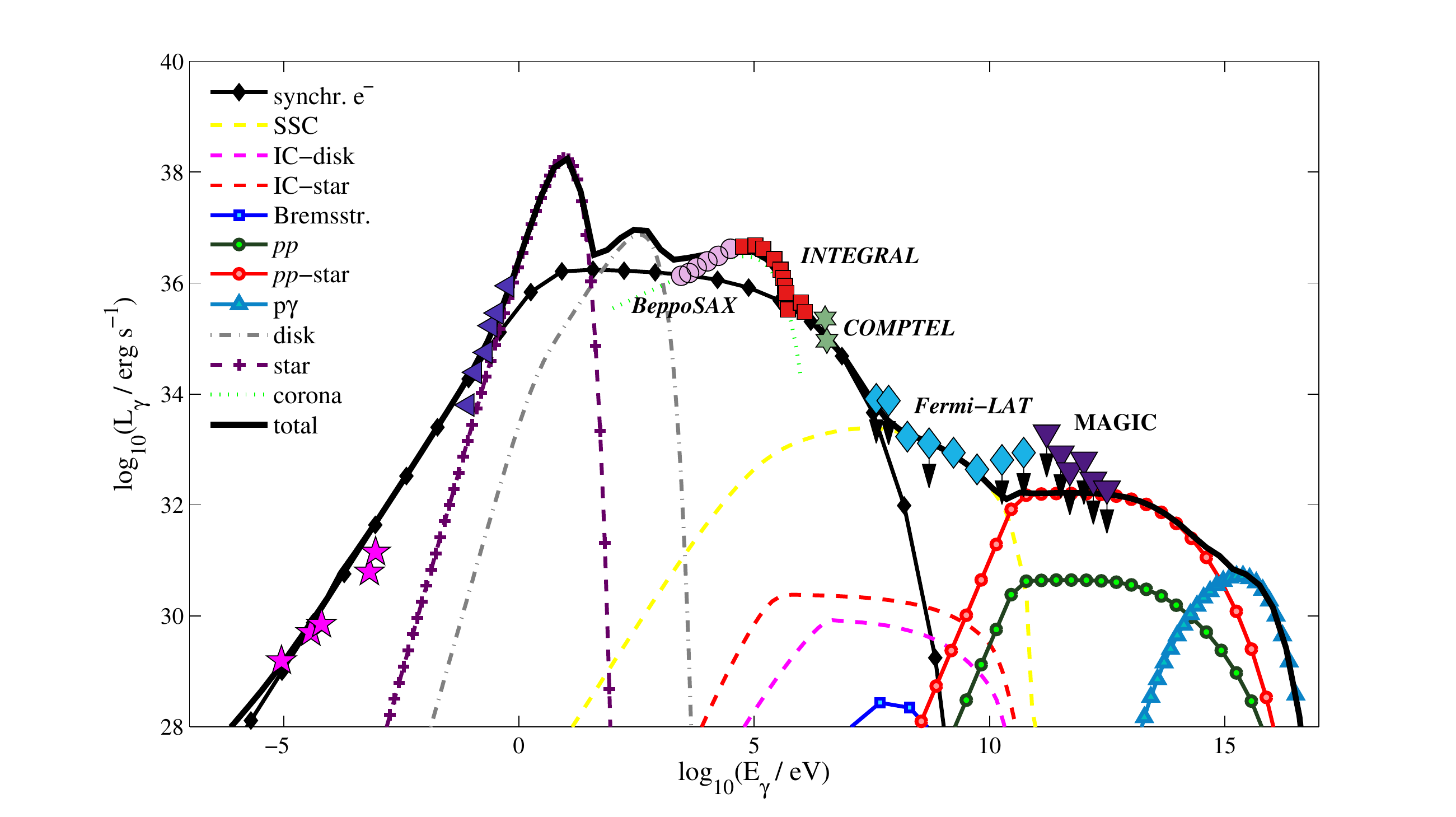}%
\caption{Spectral energy distribution of the jet of the galactic microquasar Cygnus X-1. 
From \cite{Pepeetal2015}.}%
\label{fig:Cygnus-X1}%
\end{figure}

In Fig. \ref{fig:Cygnus-X1} we show the results of the modeling of the broadband non-thermal 
emission reported by \cite{Pepeetal2015}. In this model, the emission from 1 to $\sim150$ keV 
is produced by Comptonization of disk photons in the coronae around the black hole (see, 
nevertheless, \citealt{Georganopoulosetal2002}). The synchroron radiation extends up to soft 
gamma-rays. The high-energy and very high-energy gamma-rays are the result of a combination 
of SSC, $pp$, and $p\gamma$ interactions. The dominant source of thermal protons for $pp$ is 
the wind of the star. Since these winds tend to be clumpy, flaring behaviour is expected 
around 1 TeV \citep{Romeroetal2010}.  

\subsection{Particle acceleration}
\label{breaks}

Thus due to a wide range of simultaneous broadband monitoring
campaigns and spectro-timing campaigns over the last decades, a
picture is emerging of a linked inflow/outflow system where both
``sides'' of the flow can contribute to the broadband spectrum.
Contrary to the original picture of XRBs which focused exclusively on
the accretion disk, these small accreting black holes are fully
capable of launching relativistic jets, and those jets accelerate
particles just as in blazars.  Because of their relatively higher
compactness, and thus stronger magnetic fields, jet synchrotron from
moderately high-energy particles easily reaches the X-ray band and can
compete with IC processes for the hard state spectrum.  Finally, the
detection of strong linear polarization in the soft gamma-rays from
Cyg X-1 \citep{Laurentetal2011}, and more recently the \textit{Fermi}
detection of a hard power law \citep{Zaninetal2016} seems to clinch
the question about particle acceleration in XRB jets.  Orbital
modulations hint that for this source the GeV $\gamma$-ray emission occurs further out
in the jets upon interaction with the stellar winds.  A few detections
of X-ray lines suggest hadronic content as well
\citep[e.g.,][]{MigliariFenderMendez2002,DiazTrigoetal2013}, though it
is not clear if this is entrained or directly accelerated.

The question now becomes, can we learn something new from XRBs that
will help us understand particle acceleration in jets in general, and
help reduce the large number of free parameters in the various models?
Many of the free parameters come from a lack of constraints on the jet
bulk (M/HD driven) properties, and how they couple to particle
acceleration properties (the macro-microphysics connection).  XRBs
offer a distinct advantage to AGN in this sense, given the realtime
evolution displayed in this coupling.  We will end this Section by
discussing some new explorations of this connection.

The spectral break from the flat/inverted synchrotron spectrum to a
power-law (e.g., Fig.~\ref{fig:gandhi} is the canonical signature of
the transition from self-absorbed to optically thin regimes.  The
power-law at higher frequencies reveals an accelerated particle
distribution, and the compactest part of the jet where particle
acceleration is present will dominate the emission at the break
itself.  In AGN (see Sections~\ref{structuredjets} and \ref{location})
this region is offset from the black hole, and often modeled de-facto by a
single zone where dissipation of magnetic bulk energy into particle
kinetic energy occurs via acceleration.  In XRBs, the break location
in the spectrum also corresponds to an offset from the black hole and
may correspond to a similar dissipation zone.

\begin{figure}
\centering
\includegraphics[width=0.8\columnwidth]{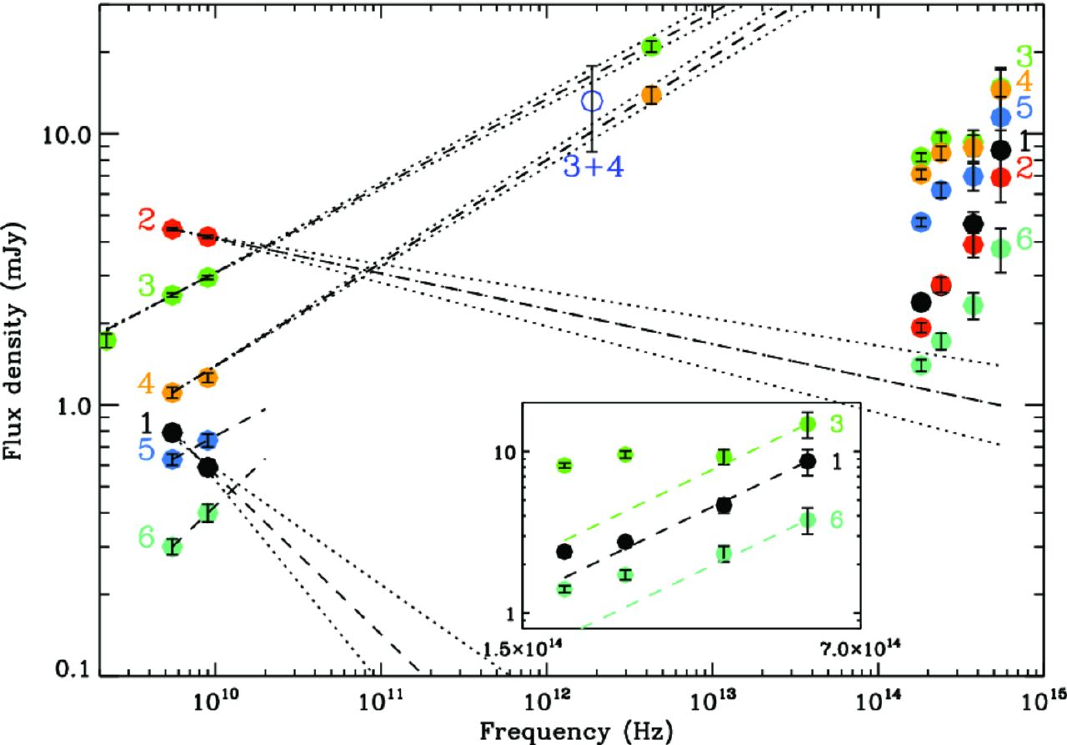}
\caption{Evolution of the radio to OIR SED during the soft to hard
  state transition as the jets reform, and become optically thick. The
  numbers (and associated colours) refer to sequential epochs, and the
  dashed lines correspond to the extrapolation of the radio spectra.
  Inset: zoom of the OIR spectra \# 1, 3 and 6.  From
  \cite{Corbeletal2013}.}
\label{fig:corbel13b}
\end{figure}

\begin{figure}
\centering
\includegraphics[width=\columnwidth]{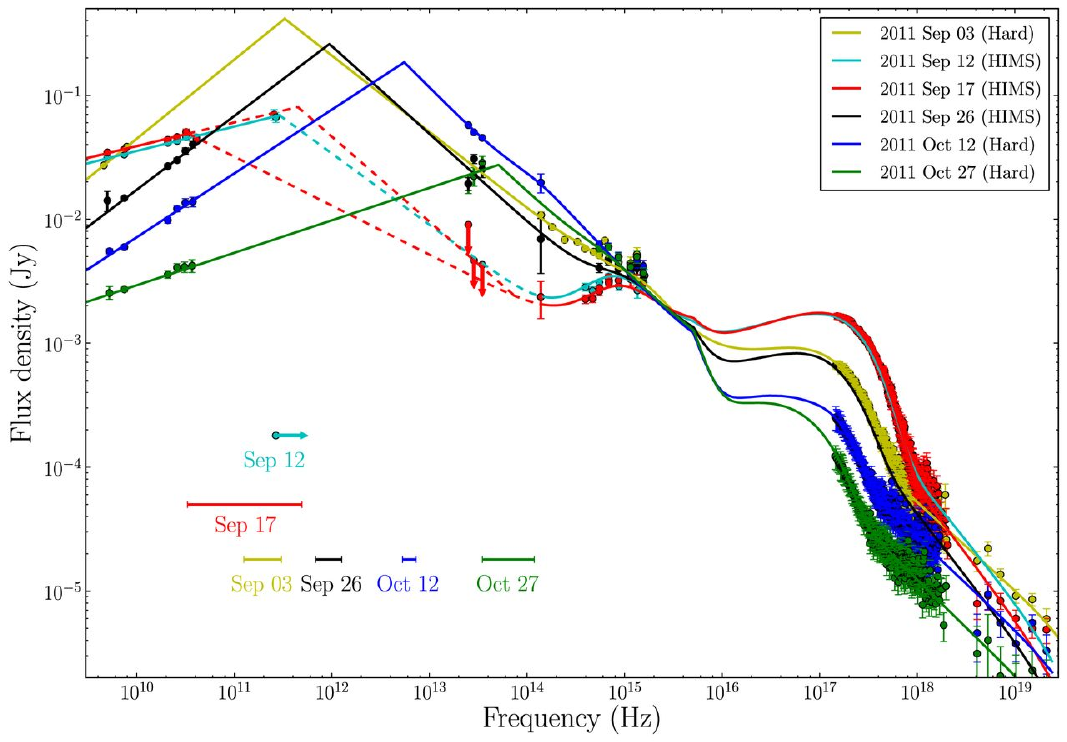}
\caption{Broadband radio to X-ray SED of MAXI J1836-194 taken during
  its 2011 outburst. Dashed lines indicate possible ranges of
  parameters, with the horizontal bars showing the uncertainty range
  for the optically thick-to-thin spectral break. Only four epochs of
  X-ray observational data are depicted to avoid crowding. The jet spectral break
  moves to higher frequency following the transition back to
  the hard state from the HIMS during a failed state transition, and
  is anticorrelated with accretion rate as indicated by the disk component.  From
  \cite{Russelletal2014}.}
\label{fig:russell14}
\end{figure}

Unlike AGN, XRB jets can barely be imaged with VLBI, however their
smaller sizes allow for much more causality between the launch point
and the outermost regions.  Assuming a conservatively low plasma
velocity of $0.3c$, the time for flow to travel from the black hole to
the outermost regions of impact on the ISM at scales on the order of a
pc \citep{Corbeletal2002} is $\sim10$ years, and most of the intrinsic
broadband SED seems to be emitted within scales $>100$ times smaller
\citep[e.g.][]{MarkoffNowakWilms2005}.  Thus an observing
campaign covering the broadband emission from radio to X-ray can
easily track the direct response of the steady jets to changes in the
inner accretion flow in real time.  However XRB jets are also seen to dismantle and reconfigure on
timescales of weeks during state changes.  During these times,
multiwavelength monitoring reveals the buildup of the accelerated
particle distribution, the transition from optically thin to
self-absorbed (see Fig.~\ref{fig:corbel13b}), and the evolution in the
spectrum including the spectral break, all as a function of changes in the accretion flow
(Fig.~\ref{fig:russell14}).

Focusing now on the spectral break, if this feature originates in the
\textit{same} region for all sources (i.e., at some fixed number of
$r_g$), then the dependence on $M_{\rm BH}$ and $\dot{m}$ can be
derived for various standard accretion scenarios \citep[see,
e.g.,][]{FalckeBiermann1995,Markoffetal2003,HeinzSunyaev2003}.  For
the case of low-luminosity sources that fit on the FP, the slope of
the correlation makes clear that these are in a radiatively
inefficient state, as explained above.  Briefly summarizing arguments
presented in \cite{HeinzSunyaev2003}, this state is gas pressure
dominated and the pressure is directly related to the particle
density, which in turn is typically defined as
$\rho=\dot{M}/(4\pi R H v)$.  Since all physical scales for black
holes can be expressed in terms of $r_g$, which is linear in mass, and
$\dot{m}\equiv \dot{M}/M_{\rm BH}$, the pressure thus depends on the
ratio $\dot{m}/M_{\rm BH}$.  MRI-driven turbulence \citep{BalbusHawley1998}
implies that the magnetic pressure is closely bound to the gas
pressure, which means $B^2\propto \dot{m}/M_{\rm BH}$ as well.  If the jet is
launched from the disk and shares the same scalings, one can solve for
the photosphere where $\tau_{\rm SSA}=1$, and show that for a fixed
plasma $\beta$ and a particle index of $p=2$, the break frequency in
the SED will scale as $\sim\dot{m}^{2/3}$, for objects of same mass
and spin.  \textit{Thus in an individual XRB, this dependence on
accretion rate can be isolated and tested}.

If the particle index $p$ is significantly steeper, or the particle
distribution is quasi-thermal, the scaling of $\dot{m}$ will be
moderated slightly to $\sim 5/8$, while if the jet were launched from
a non-advective disk (this seems ruled out by the FP, however for
completeness we include it) the scaling for all particle distributions
is $\sim 1/2$.  The point is that for any reasonable set of physical
conditions for jet launching, the internal pressure provided by the
gas and magnetic field would be expected to increase with $\dot{m}$
(for fixed spin and $M_{\rm BH}$), giving a positive scaling of the break frequency with
$\dot m$ as the jet becomes optically thick at increasingly compacter scales.  

However, already early monitoring of single sources like GX~339$-$4
indicated that the spectral break seems to move in the opposite
direction expected from purely optical depth scalings as explained
above \citep{Markoffetal2003}.  By now there are a few well-observed
outbursts from other sources that reveal the evolution of the SED
(see, e.g., Fig.~\ref{fig:russell14}).  These clearly show the
spectral break increasing in frequency as the power in the thermal
disk spectrum, and thus accretion rate, decreases.  Broader
compilations of multiple measurements from several XRBs
\cite{Russelletal2013} show a similar trend in the location of the
dissipation region.  

The current best interpretation of these campaigns is that the
location of the dissipation region that defines the break is itself
moving, likely driven by changes in the jet itself.  This
evolution can also be seen in a correlation between break frequency
and spectral index, indicating a blazar-sequence like relationship for
XRBs \citep{Koljonenetal2015}.  One important clue is that the break
seems to be relatively stable within individual sources, and from
source to source, for a given (relative) luminosity in Eddington
units.  Fits to multiple SEDs from many XRBs also show that the range
of values for the break within a given source corresponds to a range
of distance within the total length of the jets of $\sim10-10^4\;r_g$.

\begin{figure}
\centering
\includegraphics[width=\columnwidth]{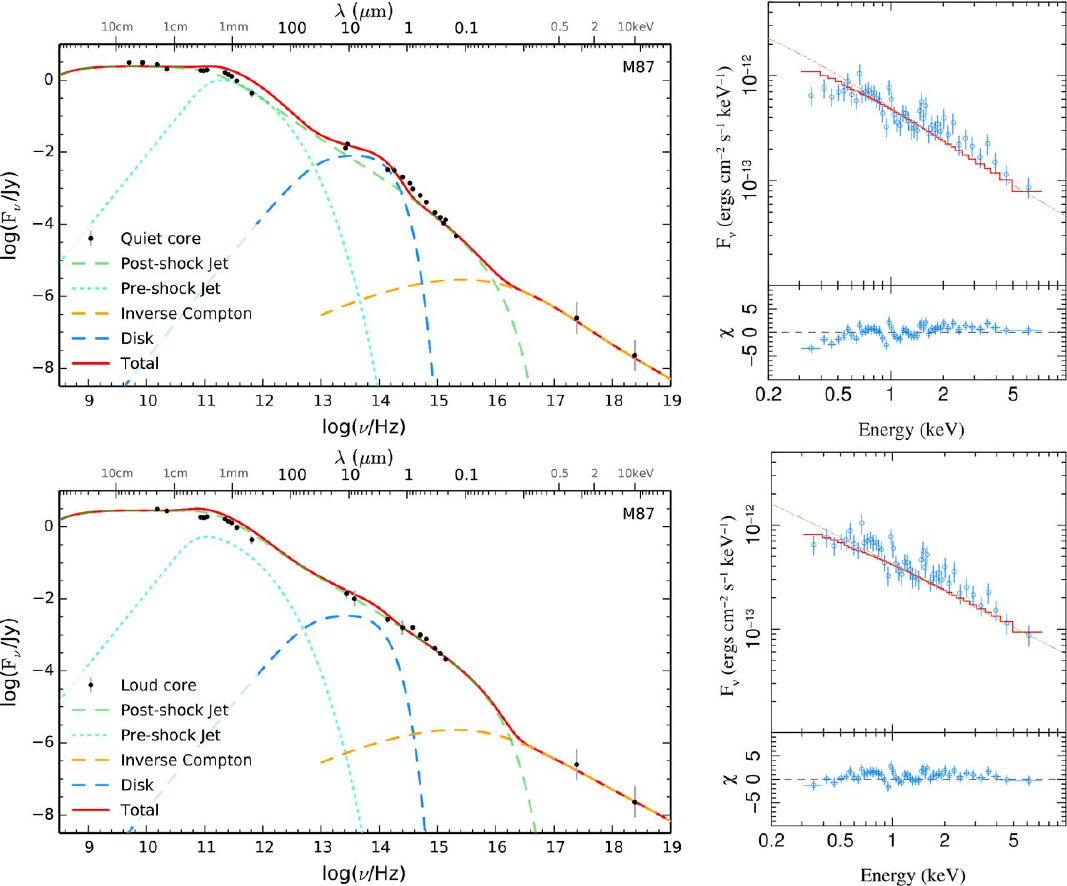}
\caption{An outflow-dominated model fit to the SED of M87, where all
  data are from within a 0.4 arcsec aperture radius (32 pc). The top
  panel shows the result for the quiescent state, the bottom panel for
  the active one. The X-ray spectrum used in the model for both states
  is the 2002 Chandra spectrum, the fit for which is shown as a
  separate plot.   The data are fit using the model in
  \cite{MarkoffNowakWilms2005} that is also applied to XRBs, and a
  location of the break implies an acceleration region of 10s of
  $r_g$, similar to XRBs as well.   From \cite{Prietoetal2016}.}
\label{fig:m87}
\end{figure}

\begin{figure}
\centering
\includegraphics[width=\columnwidth]{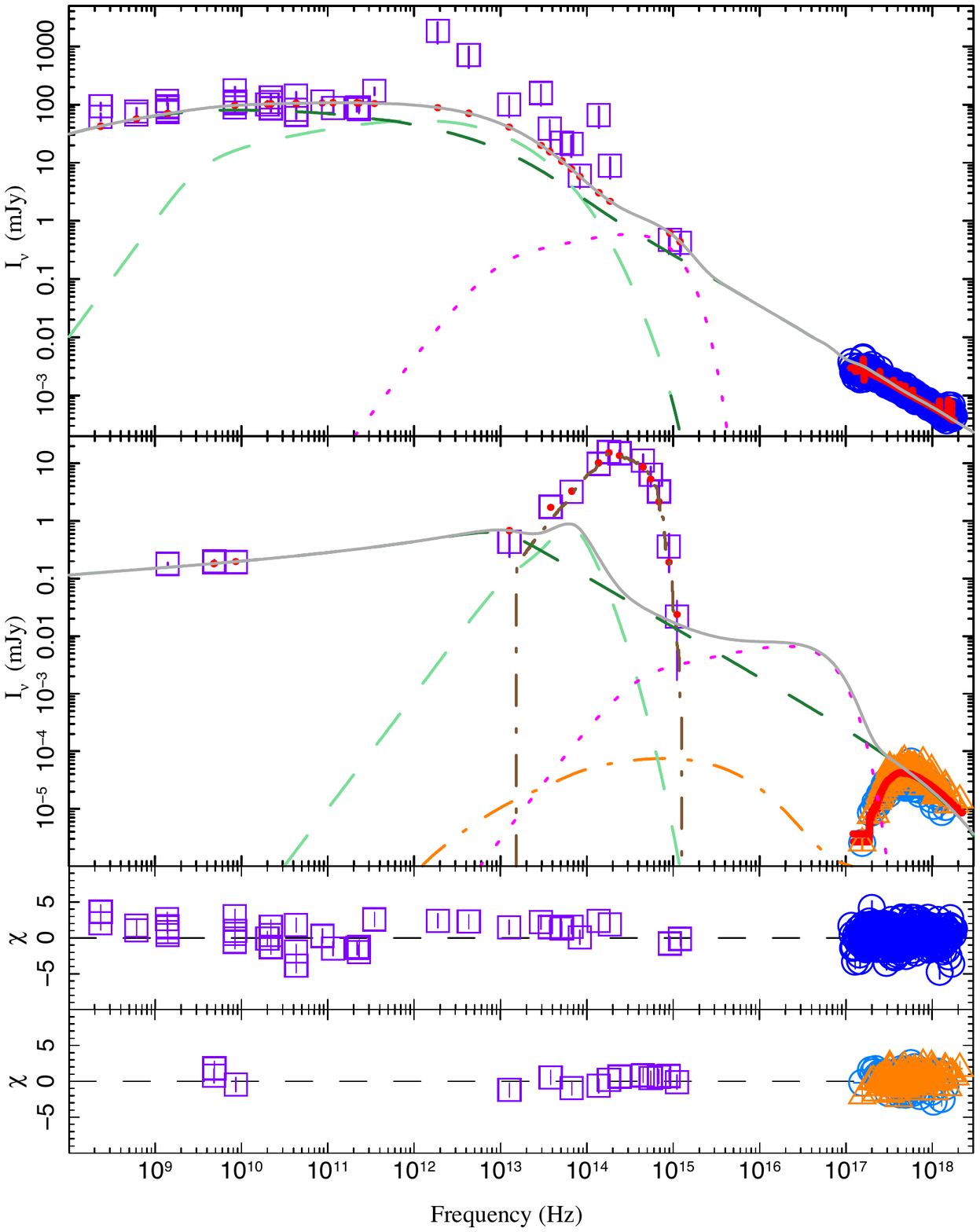}
\caption{Best joint-fit model with all physical free parameters tied
  to SEDs of similar accretion power (based on $L_{\rm X}/L_{\rm Edd}$
  from the LLAGN M81 (top) and hard state XRB V404 Cyg (bottom).
  Lines show the individual model components (light green/dashed:
  thermal synchrotron, dark green/dashed: non-thermal synchrotron,
  orange/dashed–dotted: synchrotron self-Compton, magenta/dotted:
  multicolor blackbody disk, and gray/dashed–dotted: stellar component
  in the case of V404), all absent absorption. The gray/solid line
  shows the total model, while the red/solid line and dots show the
  model after forward folding through detector space, including
  absorption.  The IR/optical part of the M81 SED also includes a
  galactic stellar and dust contribution (not fitted).   The spectral
  break in both sources implies a particle acceleration region in the jets
  starting at $300r_g$.  From \cite{Markoff2015}.}
\label{fig:markoff15}
\end{figure}

Similar to other FP properties, the location of the break seems to be
related more to the relative power (in, e.g.,$\dot m$) than the size
of the black hole.  For instance in the few hard state-like LLAGN
where quasi-simultaneous and/or high-spatial resolution data are
available, the implied break occurs within a similar range as seen in
XRBs \citep[][and see Fig.\ref{fig:m87}]{Markoffetal2008}.  In fact,
the same self-similar model (where all distances and powers are
expressed in mass-scaling units of $r_g$ and $L_{\rm Edd}$, or
$\dot m$, can be used to fit the broadband SEDs of two black holes at
opposite ends of the mass scale at similar accretion power, with all
physical scales tied (see Fig.~\ref{fig:markoff15}).  Together these
results suggest that the location of the particle acceleration zone is
a direct consequence of conditions in the base of the jets, or in
turn, the accretion flow.  This opens the door to an exploration of
whether the zone (and associated SED break) could be a predictive
property of, e.g., an MHD outflow, providing a means to drastically
reduce the number of free parameters currently employed.

In a series of papers,
\cite{PolkoMeierMarkoff2010,PolkoMeierMarkoff2013,PolkoMeierMarkoff2014}
have derived a self-similar (following on prescriptions in
\citealt{VTST2000} and \citealt{VlahakisKoenigl2003}), relativistic
MHD flow solution, including a pseudo-Newtonian treatment of gravity.
These models can be used as intermediaries between special
relativistic MHD and GRMHD simulations and observations, and also to
explore potential drivers of the correlations observed in XRBs.  For
instance, if the break location is associated with particle
acceleration at a disruption such as a recollimation shock
\citep[e.g.,][]{Marscher2008} or turbulent instabilities that form
from shear once the bulk plasma velocity exceeds the fast magnetosonic
speed, the scaling with other properties can be explored and used to
rule out various launching scenarios.  Thus the spectral breaks potentially
provide an important new pivot for connecting inflow properties to the
outflows, a longstanding issue in the field.

\subsection{Clues about particle acceleration efficiency}

XRBs have thus taught us that the location of particle acceleration
may depend on the jet bulk properties, and that this property seems to
extend to AGN, similar to the radio/X-ray correlations.  Another
indication from recent MWL campaigns is that there is a dependence of
particle acceleration efficiency on accretion power.  This phenomena
was actually first noticed in Sgr A*, where the linearly polarized
``submm bump'' synchrotron feature in the spectrum
\citep[e.g.,][]{Boweretal2015} corresponds to emission from very close
to the black hole and assumedly connects to a weak jet.  No evidence
for a spectral break or an optically thin power-law exists in the
quiescent spectrum of Sgr A*, whose synchrotron and SSC spectrum can
be explained by a mildly relativistic thermal distribution of
particles in both the corona as well as jets.  However Sgr A* also
experiences a nonthermal flaring in the X-rays approximately daily,
often accompanied by IR flaring, and models of this process suggest
the appearance of sporadic and weak particle acceleration at the level
of less than 1\% of the thermal peak
\citep[e.g.,][]{YuanQuataertNarayan2003,Dibietal2014,Dibietal2016}.

Now that there are several observations of XRBs in a similar state of
quiescence as Sgr A*, with $L_{\rm X} < 10^{-8} L_{\rm Edd}$, it seems
that the lack of a strong power law in the emitting particles may be a
universal hallmark of this state.  For instance, the SEDs of
A0620$-$00 \citep{Galloetal2007, Froningetal2011}, XTE~J1118$+$480
\citep{Plotkinetal2015}, and Swift J1357.2$-$0933
\citep{Shahbazetal2013} can all be explained with a dominant thermal
synchrotron bump, and if any power-law is present it is consistent
with being very weak and/or produced relatively far out in the jets.
The physics of this dependence of particle acceleration on accretion
rate is not yet understood, but may point to a jet power threshold
below which the structures needed to accelerate particles cannot form
or (as in the case of Sgr A*) be maintained.

One final note is that two distinct types of jets are seen in XRBs, at
different accretion powers, and also radio-quiet states are seen.
These can all occur in a single outburst of a source, thus ruling out
spin as a driver for these states.  The relationship between spin and
jet power is also being explored for XRBs, but to date the jury is
still out on this issue, due to disagreements about both how to
measure the total jet power as well as the spin
\citep[e.g.,][]{NarayanMcClintock2012,RussellGalloFender2013}.

In summary, XRBs are ideal testbeds for our understanding about black
hole accretion and jet launching in general.  Specifically they can
reveal details about how particle acceleration is linked to the bulk
plasma properties of the jets, in ways that cannot be as directly
probed in AGN due to the longer timescales.  XRB jets clearly
accelerate particles, but the same questions about leptonic vs
hadronic content still exist for these sources as for AGN.  Similar to
AGN, particularly blazars, there seems to be a
dissipation/acceleration region offset from the black hole some
distance down the jets, and a sequence in power and break frequency
that may reveal the nature of the structures providing acceleration.
The strong evidence for mass-scaling in the accretion physics around
black holes makes XRBs valuable tools for probing these questions.
The possibility that XRBs can accelerate hadrons to PeV energy is an
interesting prospect, and something that near-future instruments like
CTA will be able to test.  Their energy budgets are smaller than
supernovae, but there are many more of them, meaning that XRBs could
be an important contributing source for the Galactic CR population. 
Some models, for instance the one by \cite{Pepeetal2015}
 some of whose predictions are illustrated in Fig. \ref{fig:Cygnus-X1}, 
 can accommodate proton acceleration up to $10^{15}$ eV at the base of 
 the jet and reach $10^{16}$ eV at distances $\sim 10^{12}$ cm from the 
 compact object. These might be among the most energetic protons accelerated 
 in the Galaxy.

\section{Discussion}
\label{discussion}

Galactic and extragalactic jets present many similarities, and also many differences. 
Both seem to be powered by accretion of magnetized matter with angular momentum onto a 
compact object; they both can display apparent superluminal motions; they seem to present 
a correlation of causal significance between the accretion and ejection processes; they 
display variability along the whole electromagnetic spectrum. These and other similarities 
are often stressed. The differences, however, are also important, and frequently overlooked. 
We emphasize the following ones:

\begin{itemize}  

\item Galactic jets seem to be slower. In some cases, where the bulk motions are directly 
measured through emission lines, the plasma velocity is $\sim 0.3 c$. There are some 
exceptional objects, however, such Sco X-1, with Lorentz factors up to 10. 

\item Galactic jets become dark not far from the central source. Typical lengths are $\sim 1000$ AU.

\item They are heavy (have hadronic content) and produce thermal emission at their termination regions.            

\item They interact with stellar winds (in the case of microquasars with high-mass donor stars).

\item The size of the accretion disk is constrained by the presence of the donor star.

\item Gravitational and radiative effects of the donor star can be important to jet formation and 
propagation.

\item The disk and corona are hotter than in active galactic nuclei; magnetic fields are also higher 
close to the jet base.
\end{itemize}

These differences can yield, however, opportunities for observational and theoretical studies of physical 
situations that can be applied, with suitable modifications, to extragalactic jets. In this review we have 
seen how a relatively simple model can easily be adapted to a quantitative description of both types of 
outflows. 

Because of the proximity of Galactic jets, many physical processes that take place in mechanisms operating 
also in AGNs can be observationally probed in great detail and on
vastly shorter timescales. The use of VLBI radio observations to resolve 
jets during simultaneous X-ray and gamma-ray monitoring of microquasars remains to be fully explored. The 
implementation of this kind of multiwavelength studies would facilitate the construction of refined models 
that can be later extrapolated to AGNs and larger scales. Testing the new predictions of such models in a 
very different scenario would strengthen our understanding of the underlying physics.




\begin{acknowledgements}
  We would like to thank the ISSI staff for providing an inspiring atmosphere favourable for intense
  discussions. GER thanks support from CONICET (PIP 2014-0338) and grant AYA2016-76012-C3-1-P (Ministro de Educaci\'on, Cultura y Deporte, Spain).
  The work of M.B. is supported by the South African Research Chairs Initiative (SARChI) of the 
  Department of Science and Technology and the National Research Foundation\footnote{Any opinion,
  finding and conclusion or recommendation expressed in this material is hat of the authors, and the
  NRF does not accept any liability in this regard.} of South Africa, under SARChI Chair grant No.
  64789.
\end{acknowledgements}


\bibliographystyle{aps-nameyear}
\bibliography{romerobiblio,tavecchiobiblio,boettcherbiblio,markoffbiblio}                


\end{document}